\documentclass[aps,prb,showpacs,superscriptaddress,reprint]{revtex4-1}
\usepackage{amsmath,bm}
\usepackage{graphicx}
\graphicspath{{Plots/}}

\usepackage{soul}
\newcommand{\diag}{\mathrm{diag}\,}
\newcommand{\im}{\mathrm{Im}\,}
\newcommand{\re}{\mathrm{Re}\,}
\newcommand{\sgn}{\mathrm{sgn}\,}
\newcommand{\tr}{\mathrm{tr}\,}
\newcommand{\alphab}{\bm{\alpha}}
\newcommand{\etab}{\bm{\eta}}
\newcommand{\ii}{\mathrm{i}}
\newcommand{\hb}{\bm{h}}
\newcommand{\eb}{\bm{e}}
\newcommand{\vb}{\bm{v}}
\newcommand{\rb}{\mathbf{r}}
\newcommand{\tb}{\mathbf{t}}
\newcommand{\Ib}{\mathbf{I}}
\newcommand{\kb}{\bm{k}}
\newcommand{\vbh}{\hat{\bm{v}}}
\newcommand{\te}{\textsf{te}}
\newcommand{\tm}{\textsf{tm}}
\newcommand{\gr}{\textsf{gr}}

\newcommand{\CP}{{\textsf{CP}}}
\newcommand{\Ca}{\textsf{C}}
\usepackage{color} \definecolor{darkgreen}{rgb}{0,.5,0}
\usepackage[colorlinks,filecolor=blue,citecolor=darkgreen,unicode]{hyperref}
\usepackage[inline,nolabel]{showlabels}

\begin{document}      
\title{Casimir-Polder force and torque for anisotropic molecules close to conducting planes and their effects on CO$_2$}
\author{Mauro Antezza}
\email{mauro.antezza@umontpellier.fr}
\affiliation{Universit\'e de Montpellier Laboratoire Charles Coulomb Place Eug\`ene Bataillon - CC074 	F-34095 Montpellier Cedex 05, France}
\affiliation{Institut Universitaire de France Minist\`ere de l'\'Education Nationale, de l'Enseignement Sup\'erieur et de la Recherche 1, rue Descartes 	F-75231 Paris, France}
\author{Ignat Fialkovsky}
\email{ifialk@gmail.com}
\affiliation{CMCC, Universidade Federal do ABC, Avenida dos Estados 5001, CEP 09210-580, SP, Brazil}
\author{Nail Khusnutdinov}\email{nail.khusnutdinov@gmail.com}
\affiliation{CMCC, Universidade Federal do ABC, Avenida dos Estados 5001, CEP 09210-580, SP, Brazil}
\affiliation{Regional Scientific and Educational Mathematical Center of Kazan Federal University, Kremlevskaya 18, Kazan, Russia 420008}
\date{\today}
\begin{abstract}
We derive the Casimir-Polder force and Casimir torque expressions for an anisotropic molecule close to a conducting plane with a tensorial conductivity. We apply  our general expressions to the case of a carbon dioxide CO$_2$ molecule close to a plane with pure Hall conductivity and to graphene. We show that the equilibrium position of this linear molecule is with its main axis perpendicular to the surface. We hence conjecture a possible way to exploit the Casimir torque to mechanically improve the performance of CO$_2$ separation membranes useful for an efficient atmospheric CO$_2$ reduction.
\end{abstract}  
\maketitle 

\section{Introduction}    

The van der Waals/Casimir dispersion forces play an important role in different phenomena in physics and biology, as well as in chemistry \cite{Bordag:2009:ACE,Parsegian:2006:VdWFHBCEP,Milonni:1994:QVItQE,Woods:2016:MpoCavdWi}. The original consideration by Casimir \cite{Casimir:1948:otabtpcp} devoted to a force  between two perfect metallic plates, has been developed to include different geometries of boundaries \cite{Bordag:2009:ACE,gratingsphere,three}, non-equilibrium thermal conditions \cite{neq0,neq1,neq2,neq3,Obrecht2007}, and different materials \cite{Woods:2016:MpoCavdWi} such as graphene \cite{Bordag:2009:CibapcagdbtDm,abbas}, topological insulators \cite{Grushin:2011:tcrwtdti,*Lu:2018:cdWtfbatis,fuchs}, chiral metamaterials \cite{Zhao:2009:rCfcm} and Weyl semimetals \cite{Wilson:2015:rCfbWs,*Rodriguez-Lopez:2020:scorCitIaIIWs}.

Casimir and Polder also derived  \cite{Casimir:1948:TIrLdWf} the force between an atom in the ground state and a perfectly conducting plate taking into account the retardation of electromagnetic interactions. In a non-relativistic case, this force is the van der Waals one based on the London interaction between atoms. There exist plenty of different approaches for calculation of van der Waals/Casimir and Casimir-Polder energies and forces both at zero and non-zero temperature (see Refs.  \cite{Bordag:2009:ACE,Parsegian:2006:VdWFHBCEP,Milonni:1994:QVItQE,Woods:2016:MpoCavdWi,Intravaia2011,Obrecht2007} for details and further references). Lifshitz \cite{Lifshitz:1956:tmafbs}  was among the first who suggested a method of calculation of the Casimir-Polder force based on the consideration of vacuum fluctuations of the electromagnetic field, as well as a rarefying procedure. This approach takes into account a simple observation that for a rarefied media the dielectric permittivity $\varepsilon \approx 1 + 4\pi N \alpha$, where $N$ is the number of atoms in a unit volume and $\alpha$ is polarizability of a single atom.  

The van der Waals/Casimir and Casimir-Polder effects for anisotropic molecules are subject of intensive investigation in the recent years \cite{Babb:2005:lrasica,*Shajesh:2012:rlrfbaad,*Thiyam:2018:ddsritclt,*Buhmann:2018:cpveCPp,Thiyan:2015:acvdWCPeCO2CH4mnstf,Marachevsky:2010:CPepCSi}.  The molecule anisotropy gives life to a new phenomenon -- the Casimir torque effect which was first predicted in Refs.  \cite{Parsegian:1972:Dielectric,*Barash:1978:movdwfbab} (see also \cite{Antezza2018torque} and references therein). The Casimir energy for isotropic molecules naturally does not depend on the orientation of the latter in space, which is not the case for anisotropic molecules. The momentum of force appears which is called the Casimir torque. A molecule then is being rotated to assume a specific position in space in relation to the boundary with minima energy and zero torque.  The carbon dioxide molecule CO$_2$, for example, \cite{Thiyan:2015:acvdWCPeCO2CH4mnstf}, has an equilibrium position perpendicular to the dielectric slab, i.e. with the direction of larger polarizability being perpendicular to the surface. 

The polarizability tensor of a molecule/atom has, in general, both symmetric and antisymmetric contributions \cite{Berestetskii:1996}. The antisymmetric part is accounted for by atomic hyperfine structure \cite{Baldin:1961:Otoaan,*Manakov:1986:alf,*Becher:2018:apea,Zhizhimov:1982:PovdWf} and is defined by total angular momentum of  molecule/atom. This part of polarizability gives contribution to the $P$-odd van der Waals forces  \cite{Zhizhimov:1982:PovdWf} (see, also review \cite{Mitroy:2010:taaip}) and when interacting with a Chern-Simons surface \cite{Marachevsky:2010:CPepCSi}. 

By using the rarefying Lifshitz procedure together with the scattering matrix approach we achieve two goals in this paper: i) we obtain a general expression for Casimir-Polder force for anisotropic molecules near the conductive plane with arbitrary tensorial conductivity, and ii) we obtain a general expression for Casimir torque in this configuration. Moreover, we apply these general expressions for molecule CO$_2$ near a plane with pure Hall conductivity and in front of freestanding graphene.  The focus on studying the carbon dioxide molecules interacting with graphene originates from growing research in possible ways of implementation of graphene membranes for gas separation \cite{Huang2015} (and water desalination \cite{Cohen-Tanugi2012}) aimed to diminish the CO$_2$  footprint of the industry and to separate the gas from the atmosphere. Our predictions show that the Casimir torque tends to orient the CO$_2$  molecules in a position favoring their passage through the membrane -- facing the latter by molecules' smallest facet. Thus, possible enhancements of the effect may lead to the improvement of the efficiency of graphene-based membranes in gas separation applications.

The paper is organized in the following manner. In Sec.  \ref{Sec:IsotropicAtoms} we consider an isotropic atom near a conductive plane and implicate different forms of tensorial conductivity.  Sec. \ref{Sec:AnisotropicCase} is devoted to the derivation of the Casimir-Polder energy for anisotropic atoms/molecules in front of a conductive plane with arbitrary tensorial conductivity. We extract in a manifest form the contributions due to symmetric and antisymmetric parts of polarizability and consider some specific cases. In Sec. \ref{Sec:CPtorque} we derive the general form of Casimir torque and apply it for the CO$_2$ molecule near a surface with pure Hall conductivity and in front of graphene. We finish in Sec. \ref{Sec:Summary} with discussion and summarizing results. Some technical details of the calculations are given in Appendix \ref{Sec:Ap1}.

Throughout of the paper we use the natural units $\hbar=c=1$; the Greek indices $\mu,\nu = 1,2,3 = x,y,z$ and the Latin ones $i,j = 1,2 =x,y$. 

\section{Isotropic atoms} \label{Sec:IsotropicAtoms}

\begin{figure}[htb]
\includegraphics[width=8 cm ]{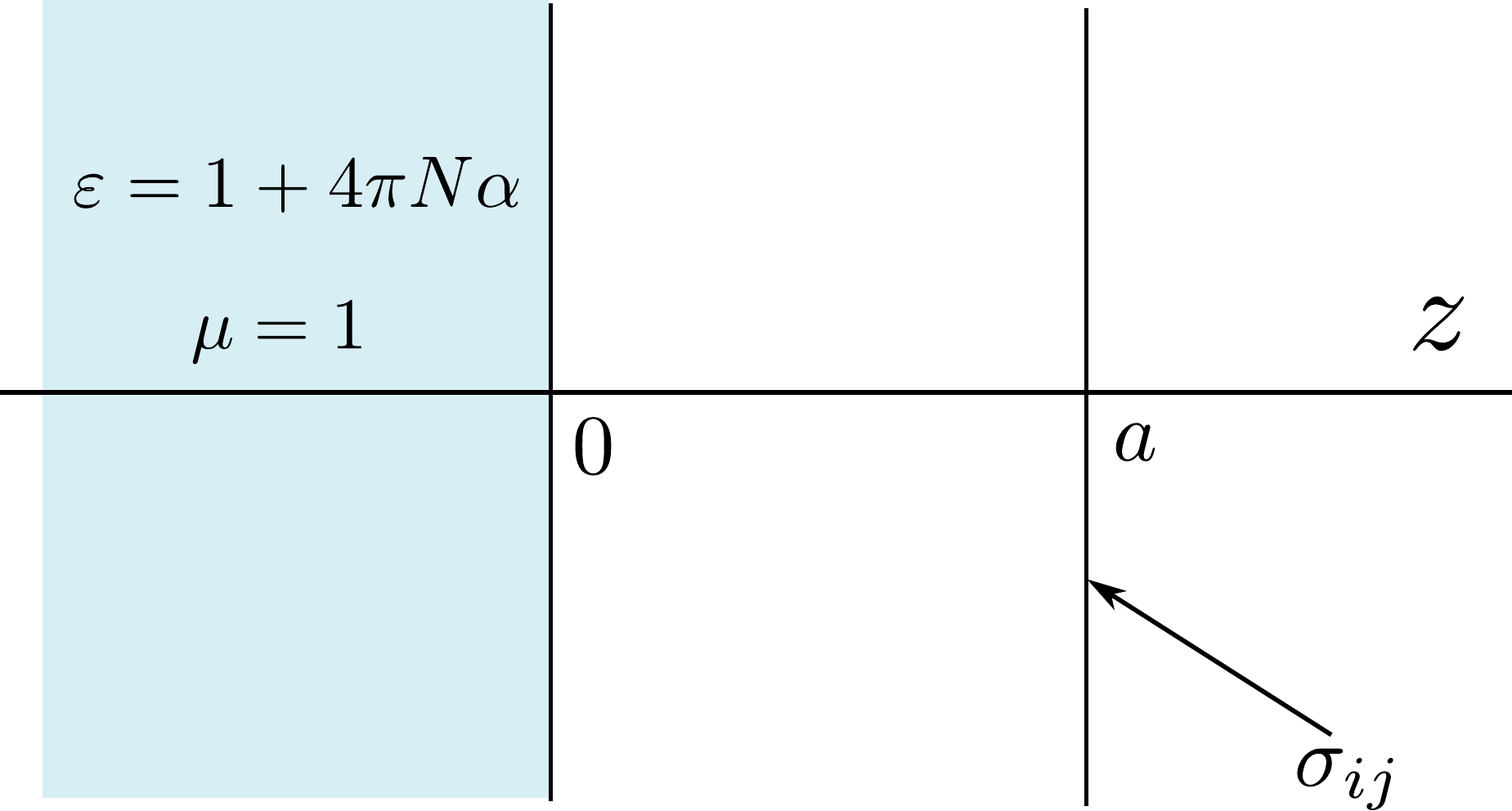}
\caption{A conductive plane  with conductivity tensor $\bm{\sigma}$ is situated at $z=a$. Semi-space $z\leq 0$ is filled by a non-magnetic dielectric. We calculate Casimir energy for this system and then rarefy dielectric, that is we set $\varepsilon = 1 + 4\pi N \alpha$ and put $N\to 0$. The CP energy is given by Eq. \eqref{eq:CP} for plane and atom at $z=0$.} \label{fig:geometry}
\end{figure}

To calculate Casimir-Polder (CP) energy we use the rarefying procedure of Lifshitz \cite{Lifshitz:1956:tmafbs}. Towards this end let us consider a conductive plane positioned perpendicular to the axis $z$ at point $z=a$ and semi-infinite non-magnetic dielectric media filling semi-space $z\leq 0$ with dielectric permittivity $\varepsilon$ (see Fig.\,\ref{fig:geometry}). Then we rarefy dielectric by considering  $\varepsilon = 1 + 4\pi N \alpha$ with $N\to 0$. The CP energy is given by relation
\begin{equation}\label{eq:CP}
\mathcal{E}_\CP = - \lim_{N\to 0} \frac{1}{N} \frac{\partial \mathcal{E}_\Ca}{\partial a}, 
\end{equation}
where $\mathcal{E}_\Ca$ is the Casimir energy calculated for the system ``conductive plane -- dielectric". 

To calculate Casimir energy $\mathcal{E}_\Ca$ for this system we use expression obtained in Ref. \cite{Jaekel:1991:Cfbptm} (see also \cite{Tse:2012:QCF}) and generalized in Ref. \cite{Fialkovsky:2018:qfcrbcs}:
\begin{equation}\label{eq:CaMat}
\mathcal{E}_\Ca =  \iint \frac{d^2k}{2(2\pi)^3} \int_{-\infty}^\infty\hspace{-1.5ex} d\xi \ln\det \left[\Ib - e^{-2 a \kappa} \rb'_0 (\ii s \kappa) \rb_a (\ii s \kappa)\right], 
\end{equation}
where $\kappa = \sqrt{\xi^2 + \kb ^2}$, $\omega = \ii \xi$, and $s = \sgn \xi$. Here $\rb_a$ and $\rb'_0$ are $2\times 2$ reflection matrices of the plane and the dielectric. 

The boundary conditions give the following expressions for reflection matrices (we denote two-dimensional tensors in bold)
\begin{eqnarray}
\rb'_0(\ii s \kappa) &=& \frac{1 - \varepsilon}{\varepsilon + 1 +  \varepsilon\frac{\kappa}{\kappa_\varepsilon} + \frac{\kappa_\varepsilon}{\kappa}} \left\{ \left(1 - \frac{\kb ^2}{\kappa^2_\varepsilon}\right)\Ib  + 2 \frac{\kb \otimes \kb }{\kappa^2_\varepsilon}\right\}, \nonumber \\
\rb_{a}(\ii s \kappa) &=& -\frac{\xi^2 \etab   + \kb  \otimes (\kb \etab  ) + \Ib \kappa |\xi | \det\etab  }{\xi^2 \tr \etab   + \kb\kb\etab  + \kappa |\xi |(1+ \det\etab  ) }, \label{eq:Ce}
\end{eqnarray}
where $\etab   = 2\pi \bm{\sigma}$, $[\kb  \otimes (\kb \etab  )]^i_j = k^i k^n \eta_{nj}$, $\kb\kb\etab = k^i k^j \eta_{ij}$,  $\kappa_\varepsilon = \sqrt{\varepsilon \xi^2 + \kb ^2}$. 

In the rarefying limit, $N\to 0$, 
\begin{equation}\label{eq:iso}
	\rb'_0(\ii s \kappa) = - \frac{\pi N \alpha}{\kappa^2} (\xi^2 \Ib + 2 \kb \otimes \kb ) + O(N^2). 
\end{equation}
Taking into account simple formula 
\begin{equation}
	\det (1+\epsilon A) = 1 + \epsilon \tr A + O(\epsilon^2),
\end{equation}
we obtain the CP energy
\begin{equation}
 \mathcal{E}_\CP = \iint \frac{d^2k}{(2\pi)^2} \int_0^\infty \frac{d\xi}{\kappa} \alpha (\ii \xi) e^{-2 a \kappa} \left(\xi^2 \tr \rb_a + 2 \kb\kb \rb_a\right),  
\end{equation}
or in the manifest form 
\begin{eqnarray}
 \mathcal{E}_\CP &=& -\iint \frac{d^2k}{(2\pi)^2} \int_0^\infty \frac{d\xi}{\kappa} \alpha (\ii \xi) e^{-2 a \kappa} \nonumber \\
 &\times& \frac{\xi^4 \tr \etab   + \kb\kb\etab(\xi^2 + 2 \kappa^2) + 2 \kappa^3 \xi \det\etab  }{\xi^2 \tr \etab   + \kb\kb\etab  + \kappa \xi (1+ \det\etab  ) }.  \label{eq:Iso}
\end{eqnarray}
This expression is valid for an arbitrary form of the conductivity. For a plane with ideal conductivity, $\etab \to \infty$, the CP energy tends to that for the ideal plane, $\mathcal{E}_\CP \to	\mathcal{E}_\CP^{id}$, where 
\begin{equation}
\mathcal{E}_\CP^{id} = - \frac{1}{16\pi a^4} \int_0^\infty dz\, e^{-z} \alpha\left(\frac{\ii z}{2a}\right)(2 + 2 z + z^2). \label{eq:EcpIdeal}
\end{equation}

For large distances between the atom and the plane we change integrand variables $\kb\to \kb/a, \xi \to \xi/a $ and take the limit $a\to \infty$.  We observe that the CP energy is proportional to that for perfect metal, $\mathcal{E}_\CP^{\infty} = -3\alpha(0)/8\pi a^4$, with coefficient depending on the conductivity of the plane. This was observed in Ref.  \cite{Khusnutdinov:2016:CPefasocp} for isotropic conductivity.  Note that the actual parameter of expansion depends on the structure of the polarization tensor $\alpha$, which must include dimensionful parameters describing the properties of the atom/molecule, see \eqref{eq:a D}.

Let us consider the general form of the conductivity tensor,
\begin{equation}\label{eq:etaGen}
	\etab   = X \Ib + Y \frac{\kb \otimes \kb }{\kb ^2} + Z \bm{\epsilon}, 
\end{equation}
where all coefficients are functions of $(\xi,k)$ and $\bm{\epsilon}$ is the totally  antisymmetric tensor. In particular,  the graphene conductivity tensor has such  structure \cite{Bordag:2009:CibapcagdbtDm,*Fialkovsky:2011:FCefg,*Bordag:2016:ECefdg} with $X=\eta_\te, Y = \eta_\tm   - \eta_\te$ and $Z =0$ \cite{Khusnutdinov:2019:lteotcfefaaiwacp}. Here $\eta_{\tm ,\te}$ is conductivity of TM and TE modes, correspondingly. If $Z\not = 0$, the parity anomaly is present \cite{Fialkovsky:2018:qfcrbcs}. The pure Hall conductivity corresponds to the case with $X=Y=0$ and $Z \not = 0$. 
  
Taking the polar coordinates for $(k_1,k_2)$ and integrating over angular variable we obtain that 
\begin{eqnarray}
	\mathcal{E}_\CP &=& -\int\!\!\!\!\!\int_0^\infty \frac{kdk d\xi}{2\pi\kappa} \frac{\alpha (\ii \xi)}{W } e^{-2 a \kappa}\left\{\kappa ^2 \left(2 \kappa ^2-\xi ^2\right) (X+Y)  \right. \nonumber \\
	&+& \left. 2 \kappa ^3 \xi  \left(X^2+X Y+Z^2\right)+\xi ^4 X\right\},  \label{eq:Iso1}
\end{eqnarray}
where 
\begin{equation}
	W  = \kappa ^2 (X+Y)+ \kappa  \xi  \left(X^2+X Y+Z^2+1\right)+X \xi ^2.
	\label{eq:W}
\end{equation}

For isotropic conductivity $\etab   = X\, \Ib$ the CP energy, 
\begin{eqnarray}
 \mathcal{E}_\CP &=& -\int_0^\infty \frac{dz z^3}{2\pi} \int_0^1 \hspace{-1ex}dx\, \alpha (\ii xz) e^{-2 z  a}\nonumber \\
 &\times& \left\{\frac{X x^3}{X x + 1} + \frac{X (2 -x^2)}{X  + x}\right\},
\end{eqnarray}
coincides with expression obtained in Ref. \cite{Khusnutdinov:2016:CPefasocp}. Here $X = X (xz)$.

Let us consider pure Hall conductivity $X=Y = 0$ and $Z\not = 0$. For constant $Z$ one has 
\begin{equation}
 \mathcal{E}_\CP = \frac{Z^2}{1+Z^2} \mathcal{E}_\CP^{id},  
\end{equation}
where the CP energy for the plane with ideal conductivity, $\mathcal{E}_\CP^{id}$, is given by Eq. \eqref{eq:EcpIdeal}. The same result was obtained in Refs. \cite{Marachevsky:2010:CPepCSi,*Buhmann:2018:cpveCPp} for CP interaction of an atom with a Chern-Simons plane. The parameter of Chern-Simons interaction in those references coincides exactly with the Hall conductivity $Z$ in the above formulas. 

For a two-dimensional electron systems in a strong magnetic field $B$ we have $\sigma_{xx} \ll \sigma_{xy}$ (see Ref. \cite{Tse:2012:QCF}) and 
\begin{equation}
Z = 2\pi \sigma_{xy} = \nu \alpha_{\rm QED},
\end{equation}
where $\nu = 2\pi n /(eB)$ is the Landau-level filling factor and $\alpha_{\rm QED}$ is the fine structure constant. The leading contribution over $\alpha_{\rm QED} \ll 1$ reads
\begin{equation}
 \mathcal{E}_\CP = Z^2 \mathcal{E}_\CP^{id} = \alpha^2_{\rm QED} \nu^2 \mathcal{E}_\CP^{id}.   
\end{equation}
Therefore, the CP energy is quantized, too,  alongside with Casimir energy \cite{Tse:2012:QCF} and suppressed by the same factor $\alpha^2_{\rm QED}$. The CP force is always attractive, for any distance and any value of the Hall conductivity. The detailed calculations for graphene in the strong magnetic field have already been done in Ref. \cite{Cysne:2014:TtCivmeig}.

In the general case with a combination of the isotropic and the Hall conductivities we set $Y=0$ and have 
\begin{equation}
\etab =X \Ib + Z\bm{\epsilon}.
\end{equation}
The CP energy reads
\begin{eqnarray}
 \frac{\mathcal{E}_\CP}{\mathcal{E}_\CP^{\infty}} &=& \frac{4}{3}\int_0^\infty dy y^3 \int_0^1 dx \frac{\alpha (\ii x y/a)}{\alpha(0)} e^{-2 y} \nonumber \\ 
 &\times& \frac{(2-x^2+x^4)X + 2x (X^2 + Z^2)}{(1+x^2) X + x (1 + X^2 +Z^2)},  
\end{eqnarray}
where $\mathcal{E}_\CP^{\infty} = -3 \alpha(0)/8\pi a^4$ is the asymptotic of the Casimir-Polder energy for ideal plane \eqref{eq:EcpIdeal} and the arguments of $X$ and $Y$, $(\xi,k) \to (yx/a, y\sqrt{x^2-1}/a)$. The function $2-x^2+x^4$ is always positive in domain of integration and therefore the CP energy is always negative and the CP force is attractive for any relation between $X$ and $Z$.  

\section{Anisotropic case} \label{Sec:AnisotropicCase}

To include in the consideration an anisotropic dynamic  polarizability we consider a matter in semi-space with tensorial dielectric permittivity $\varepsilon_{\mu\nu}$. First of all we have to consider the scattering problem for an anisotropic semi-space, $z \leq 0$ and vacuum for $z>0$  and then use the rarefying Lifshitz  procedure: $\varepsilon_{\mu\nu} = \delta_{\mu\nu} + 4\pi N \alpha_{\mu\nu} $ with $N\to 0$. 

In general, the polarizability tensor $\alpha_{\mu\nu} (\omega)$ may be expressed in terms of dipole matrix elements \cite{Berestetskii:1996}, 
\begin{equation}
\alpha_{\mu\nu} = \sum_n \left(\frac{\mathrm{D}_{\mu\nu}^n}{\omega_{n0} - \omega - \ii 0} + \frac{\mathrm{D}_{\nu\mu}^n}{\omega_{n0} + \omega + \ii 0}\right), 
\label{eq:a D}
\end{equation}
where $\omega_{n0} = \omega_n - \omega_0$, and $\mathrm{D}_{\mu\nu}^n = \langle 0| \hat d_\mu | n \rangle \langle n| \hat d_\nu | 0 \rangle $. The scattering tensor has the same form but with $-\ii 0$ in the second term \cite{Berestetskii:1996}. For the case of elastic scattering \cite{Berestetskii:1996}, $\omega \not = \omega_{n0}$ it is possible to omit $\ii 0$ in denominator and  the polarizability tensor becomes Hermitian. Note, that such approximation, strictly speaking, violates the Kramers-Kronig relations. The polarizability tensor may be divided into symmetric and antisymmetric parts $\alpha_{\mu\nu}  = \alpha^s_{\mu\nu} + \alpha^a_{\mu\nu}$ and for elastic scattering case we have \cite{Marachevsky:2010:CPepCSi,*Buhmann:2018:cpveCPp}, 
\begin{eqnarray}
\alpha^s_{\mu\nu} &= &\frac{\alpha_{\mu\nu}  + \alpha_{\nu\mu}}{2} = \re \alpha_{\mu\nu}  = \sum_n \frac{2\omega_{n0} \re \mathrm{D}_{\mu\nu}^n}{\omega_{n0}^2 - \omega^2}, \nonumber \\
\alpha^a_{\mu\nu} &= & \frac{\alpha_{\mu\nu}  - \alpha_{\nu\mu}}{2}  = \ii\, \im \alpha_{\mu\nu}  = \sum_n \frac{2\ii \omega \im \mathrm{D}_{\mu\nu}^n}{\omega_{n0}^2 - \omega^2},\label{eq:SA}
\end{eqnarray}
and for the negative frequencies $\omega <0$ we use the relation \cite{Berestetskii:1996}: $\alpha_{\mu\nu} (\omega) = \alpha_{\mu\nu} (-\omega)$. At the imaginary axes, $\omega = \ii \xi$ it becomes real. 

It was shown \cite{Baldin:1961:Otoaan,*Manakov:1986:alf,*Becher:2018:apea,Zhizhimov:1982:PovdWf} that taking into account
the atomic hyperfine structure, the polarizability tensor maybe decomposed in three contributions -- scalar (diagonal), vector (antisymmetric), and tensor (symmetric and traceless). The sum of the scalar and tensor parts correspond to the symmetric part, $\alpha^s_{\mu\nu}$, and the vector one -- to the antisymmetric part, $\alpha^a_{\mu\nu}$, of polarizability \eqref{eq:SA}. The antisymmetric part is defined by the total angular momentum, $\mathbf{J}$, of an atom \cite{Zhizhimov:1982:PovdWf,Le:2013:dpaalfgtac}:
\begin{equation}\label{eq:av}
	\alpha^a_{\mu\nu}(\omega ) = \ii \alpha_v (\omega )\epsilon_{\mu\nu\rho} J^\rho,
\end{equation} 
where $\epsilon_{\mu\nu\rho}$ is completely antisymmetric tensor. The vector polarizability gives contribution to the P-odd van der Waals forces  \cite{Zhizhimov:1982:PovdWf} (see, also review \cite{Mitroy:2010:taaip}). 

The CP energy is obtained by the same rule \eqref{eq:CP}  (see Appendix \ref{Sec:Ap1} and Ref. \cite{Emelianova:2020:Cedapsc}). The CP energy reads 
\begin{eqnarray*}
\mathcal{E}_\CP &=& \iint  \frac{d^2k}{4\pi^2}\int_{0}^{\infty}\!\! \frac{d\xi}{\kappa}  e^{-2 \kappa  a} r^{ij}_a \\ 
&\times&\left[\xi^2 \alpha_{ji} +  \alpha_{ni} k^n k_j + \alpha_{33} k_ik_j\right],  
\end{eqnarray*}
where $\alpha_{ij} = \alpha_{ij}(\ii \xi)$ and $r^{ij}_a$ is the reflection matrix of the plane at $z=a$.   

For a plane with arbitrary tensorial conductivity, the matrix $\rb_a$ has form given by Eq. \eqref{eq:Ce} and the CP energy becomes
\begin{eqnarray}
\mathcal{E}_\CP(\alpha) &=& -\iint  \frac{d^2k}{(2\pi)^2}\int_0^{\infty} \frac{d\xi}{\kappa}  \nonumber \\ 
&\times& \frac{e^{-2 \kappa  a}}{\xi^2 \tr \etab   + \kb\kb\etab  + \kappa \xi (1+ \det\etab  )} \nonumber \\ 
&\times&\left\{\xi^2 \left[\xi^2 \tr(\etab \alphab) +  \kb  \kb \left(\etab\alphab +  \alphab\etab\right) \right]\right.\nonumber \\ 
&+&\left. (\kb\kb\etab) \left[( \kb \kb\alphab) + \alpha_{33} \kappa^2\right]\right.\nonumber \\ 
&+& \left.\kappa \xi \det \etab  \left[\xi^2 \tr\alphab + ( \kb\kb\alphab) + \alpha_{33}  \kb ^2 \right] \right\}.\label{eq:cas}
\end{eqnarray}
Here, $( \kb  \kb (\etab\alphab)) = k^i k^j \eta_{in}\alpha_{nj}$ and $( \kb  \kb (\alphab\etab)) = k^i k^j \alpha_{in}\eta_{nj}$. We observe from this expression that the result does not depend on the components $\alpha_{3i}$ and $\alpha_{i3}$. This point was observed in Refs. \cite{Marachevsky:2010:CPepCSi,*Buhmann:2018:cpveCPp}. We may rewrite above expression in an invariant form by changing $\alpha_{33} \to \alpha_{\mu\nu} n^\mu n^\nu$, where vector $n^\mu$ is the unit vector perpendicular to the plane and assuming other $2D$ tensors living in this plane.   

According with the partition \eqref{eq:SA} we can represent the energy as sum of two contributions $\mathcal{E}_\CP = \mathcal{E}_\CP^s + \mathcal{E}_\CP^a$ where $\mathcal{E}_\CP^s = \mathcal{E}_\CP(\alpha^s)$, $\mathcal{E}_\CP^a = \mathcal{E}_\CP(\alpha^a)$, and 
\begin{equation}
\mathcal{E}_\CP^a = -\iint  \frac{d^2k}{(2\pi)^2}\int_0^{\infty}\!\!  \frac{\kappa \xi^2  \tr(\etab \alphab^a)e^{-2 \kappa  a} d\xi}{\xi^2 \tr \etab   + \kb\kb\etab  + \kappa \xi (1+ \det\etab  )}, \label{eq:Asym}
\end{equation}
is contribution of the antisymmetric part of polarizability. In manifest form $\tr(\etab \alphab^a) = -\alpha^a_{12}(\eta_{12}-\eta_{21}) = \widetilde{\alpha}_v(\xi)J_z (\eta_{12}-\eta_{21})$, where $\ii \alpha_v(\ii \xi) = -\widetilde{\alpha}_v(\xi)$. This contribution is zero either for symmetric polarizability $\alphab$, or for the symmetric conductivity $\etab$. 

In the case of isotropic polarizability, $\alpha_{\mu\nu} = \alpha \delta_{\mu\nu}$ we obtain Eq. \eqref{eq:Iso}.  In the case of anisotropic polarizability we can take a formal limit $\etab \to \infty$ to consider perfect metal plate, obtaining
\begin{eqnarray}
	\mathcal{E}_\CP^{id} &=& - \frac{1}{32\pi a^4} \int_0^\infty dz e^{-z}\left\{ \alpha^\mu_\mu (1+z + z^2) \right.\nonumber \\
	&+& \left. \alpha_{33} (1 + z - z^2) \right\}, \label{eq:CPid}
\end{eqnarray}
where $\alpha_{\mu\nu} = \alpha_{\mu\nu} (\ii z/2a)$. There is no contribution from the off-diagonal components. This expression is a generalization the CP energy for the isotropic molecule \eqref{eq:EcpIdeal}. For large distance, $a\to \infty$, we obtain 
\begin{equation}\label{eq:Eid}
\left. \mathcal{E}_\CP^{id} \right|_{a\to \infty} = \mathcal{E}_\CP^{\infty} = - \frac{\alpha^\mu_\mu(0)}{8\pi a^4}, 
\end{equation}
in accordance with Refs. \cite{Marachevsky:2010:CPepCSi,Thiyan:2015:acvdWCPeCO2CH4mnstf}. It depends on the trace of the total polarizability tensor, and therefore far from the surface, the CP energy does not depend on the orientation of the molecule.  

For conductivity tensor in the form \eqref{eq:etaGen} we obtain 
\begin{eqnarray}
	\mathcal{E}_\CP^s &=& -\iint _0^\infty \frac{k dk d\xi}{4\pi \kappa} e^{-2 \kappa  a}  \frac{(\alpha_{11} + \alpha_{22}) a^{11} + \alpha_{33} a^{33}}{W },  \nonumber \\
	\mathcal{E}_\CP^a &=& \iint _0^\infty \frac{k dk d\xi}{\pi } e^{-2 \kappa  a}  \frac{Z\kappa \xi^2 \alpha^a_{12}}{W },\label{eq:Es}
\end{eqnarray}
where  $W$ is given in \eqref{eq:W} and
\begin{eqnarray}
a^{11} &=& \kappa ^4 (X+Y) + \kappa  \xi  \left(\kappa ^2+\xi ^2\right) \left(X^2+X Y+Z^2\right)\nonumber\\ 
&+&\xi ^4 X, \nonumber \\
a^{33} &=& 2 \kappa k^2 \left(\kappa  (X+Y) + \xi  \left(X^2+X Y+Z^2\right)\right).
\end{eqnarray}

For pure constant Hall conductivity $X=Y=0$ one has 
\begin{eqnarray*}
	\mathcal{E}_\CP &=& \frac{Z^2}{1+Z^2} \mathcal{E}_\CP^{id}  \\
	&+&  \frac{Z}{4\pi a^4 (1+Z^2)} \int_0^\infty dz e^{-z}  \alpha^a_{12}\left(\frac{\ii z}{2a}\right) z(1+2z), 
\end{eqnarray*}
where $\mathcal{E}_\CP^{id}$ is given by Eq. \eqref{eq:CPid}. The diagonal part of polarizability gives the contribution which is even with respect to conductivity $Z$, while the off-diagonal part is odd and it is zero for symmetric polarizability. The same result was obtained in Refs. \cite{Marachevsky:2010:CPepCSi,*Buhmann:2018:cpveCPp} for CP interaction atom with Chern-Simons plane. 

\section{Casimir torque} \label{Sec:CPtorque}

\begin{figure}[htb]
	\includegraphics[width=8 cm ]{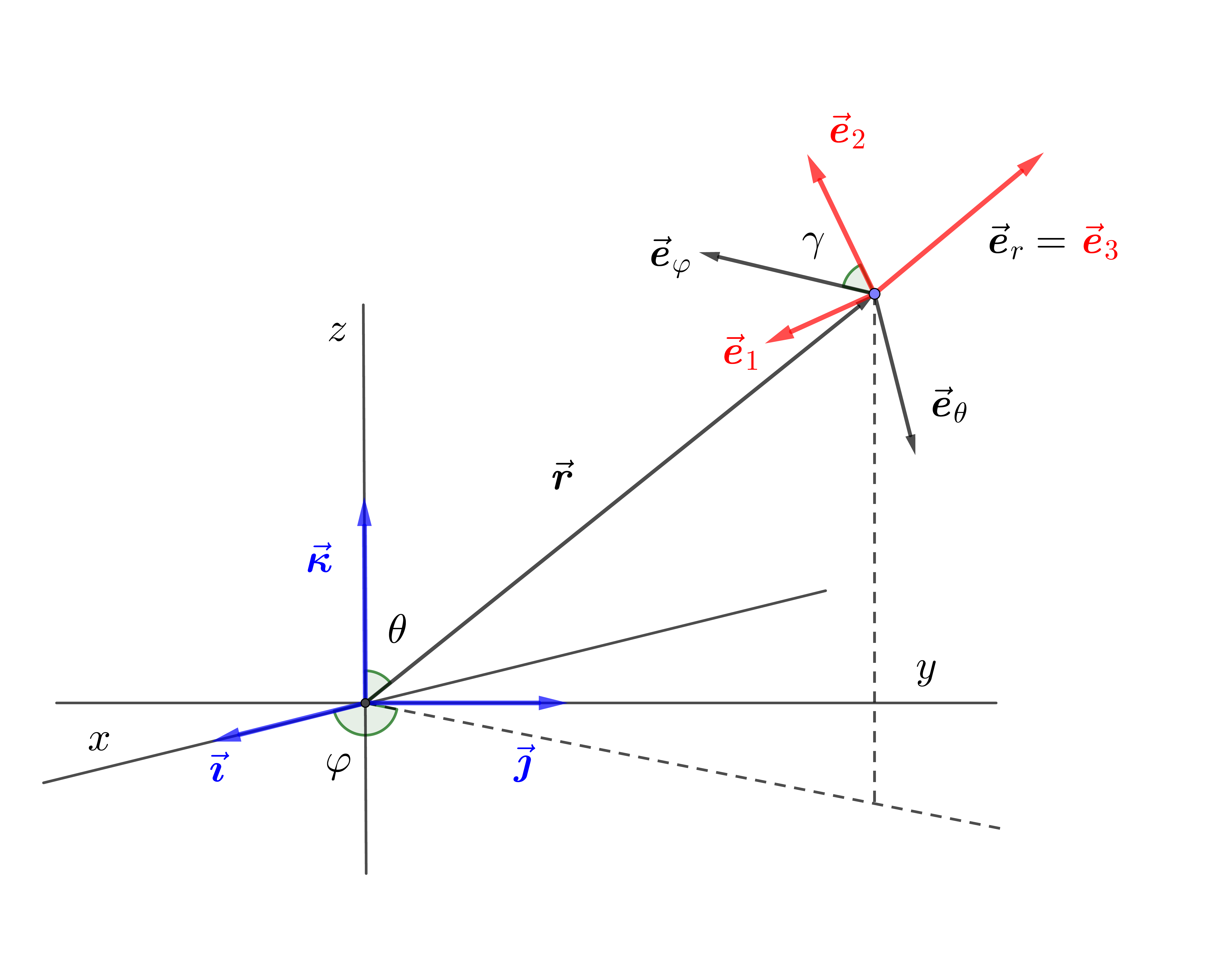}
	\caption{The conductive plane is $(x,y)$, with a tensor conductivity $\bm{\sigma}$. The eigenvectors of the polarizability tensor of a  molecule are $(\vec{e}_1, \vec{e}_2, \vec{e}_3)$ which are obtained by rotation the orthonormal spherical basis $(\vec{e}_\theta, \vec{e}_\varphi, \vec{e}_r)$ on the angle $\gamma$ around $\vec{e}_r$.} \label{fig:sis}
\end{figure}

To calculate the Casimir torque we have to characterize orientation  in space the anisotropically polarizable molecule. In the Eq. \eqref{eq:cas} a molecule is situated in such way that it has polarizability tensor $\alpha_{\mu\nu}$ in the Cartesian coordinates $(x,y,z)$. Let us consider symmetric tensor of the atomic polarizability, only. Suppose that the principal axes of the molecule are defined by the eigenvectors $(\vec{e }_1, \vec{e}_2, \vec{e}_3)$ which may be obtained by rotation the spherical basis $(\vec{e}_\theta, \vec{e}_\varphi, \vec{e}_r)$ on the angle $\gamma$ around $\vec{e}_r$ \cite{Thiyan:2015:acvdWCPeCO2CH4mnstf} (see Fig.\,\ref{fig:sis}),  
\begin{eqnarray}
\vec{e}_1 &=&  \vec{e}_\theta \cos\gamma - \vec{e}_\varphi \sin \gamma , \nonumber\\
\vec{e}_2 &=& \vec{e}_\theta \sin\gamma + \vec{e}_\varphi \cos \gamma , \nonumber\\
\vec{e}_3 &=& \vec{e}_r. \label{eq:basis}
\end{eqnarray}
Note that setting $\vec{e}_3 = \vec{e}_r$ is not restrictive as we can always choose freely the origin of the coordinate system.   

The matrix $\mathbf{T}$ of transformation from the old basis $(\vec{\imath},\vec{\jmath},\vec{k })$ to the new one $(\vec{e}_1,\vec{e}_2,\vec{e}_3)$ is defined by the relation 
\begin{equation}
(\vec{e}_1\ \vec{e}_2\ \vec{e}_3)= (\vec{\imath}\ \vec{\jmath}\ \vec{k })  \cdot \mathbf{T}.
\end{equation}
The matrix $\mathbf{T}$  may easily be found from the relations \eqref{eq:basis}. In a new basis, the polarization tensor has a diagonal form 
\begin{equation}
[\alpha_{\mu\nu}] =  \diag (\alpha_{11}, \alpha_{22}, \alpha_{33}) = \mathbf{T}^T [\alpha'_{\mu\nu}] \mathbf{T}.
\end{equation}
Therefore,  to calculate the energy we use the following representation of polarization tensor $[\alpha'_{\mu\nu}] = \mathbf{T}[\alpha_{\mu\nu}] \mathbf{T}^T$.  

For $\theta = \varphi = \gamma =0$, i.e when the principal axes of molecule are oriented along the Cartesian axes $(x,y,z)$, the old basis $(\vec{\imath},\vec{\jmath},\vec{k})$ coincides with new one  $(\vec{e}_1,\vec{e}_2,\vec{e}_3)$ and the CP energy reads
\begin{equation}\label{eq:E0}
	\mathcal{E}_\CP^0 = -\int\!\!\!\!\int  \frac{d^2k}{(2\pi)^2}\int_0^{\infty} \frac{d\xi}{\kappa}  \frac{e^{-2 \kappa  a} \alpha_{\mu\nu} A^{\mu\nu}}{\xi^2 \tr \etab   + \kb\kb\etab  + \kappa \xi (1+ \det\etab  )},
\end{equation}
where
\begin{eqnarray}
A^{11} &=& (\kb\kb\etab+ \kappa \xi \det \etab + \xi^2 \tr \etab)(\xi^2+k_1^2) - \xi^2 \kappa^2 \eta_{22}, \nonumber \\
A^{22} &=& (\kb\kb\etab + \kappa \xi \det \etab + \xi^2 \tr \etab)(\xi^2+k_2^2) - \xi^2 \kappa^2 \eta_{11}, \nonumber \\
A^{33} &=& (\kappa\, \kb\kb\etab + \xi \kb ^2\det \etab) \kappa.
\end{eqnarray}

In general case, $\theta \not = 0, \varphi \not = 0, \gamma \not =0$, and we obtain the CP energy 
\begin{equation}
\mathcal{E}_\CP = \mathcal{E}_\CP^0 + \Delta \mathcal{E}_\CP, \label{eq:EcpAngles}
\end{equation} 
where
\begin{eqnarray}
	\Delta \mathcal{E}_\CP &=& -\iint  \frac{d^2k}{(2\pi)^2}\int_0^{\infty} \frac{d\xi}{2\kappa}  \nonumber \\ 
	&\times& \frac{e^{-2 \kappa  a}}{\xi^2 \tr \etab + \kb\kb\etab  + \kappa \xi (1+ \det\etab  )}\nonumber \\
	&\times&\left\{ (\alpha_{11}-\alpha_{22}) S_{12} +  (\alpha_{22}-\alpha_{33}) S_{23}\right\}.
\end{eqnarray}
Here 
\begin{eqnarray}
S_{12} &=& \cos 2\gamma \left(B_{12} \sin 2\varphi + B_{11} \cos 2\varphi\right)\nonumber \allowdisplaybreaks \\
&-& \cos\theta \sin 2\gamma \left(B_{12} \cos 2\varphi + B_{11} \sin 2\varphi \right)\nonumber \allowdisplaybreaks \\
&-& B_{11} + \cos^2 \gamma S_{23} ,\nonumber \allowdisplaybreaks \\
S_{23} &=&  \sin^2\theta \left\{ k^2 (\kb\kb\etab + \kappa \xi \det \etab + \xi^2 \tr \etab) \right.\nonumber \allowdisplaybreaks\\
&-& \left. \xi^2 \kappa (2\xi \det\etab +\kappa \tr\etab)\right. \nonumber \allowdisplaybreaks \\
&-& \left. B_{12} \sin 2\varphi  - B_{11}\cos 2\varphi \right\},
\end{eqnarray}
and
\begin{eqnarray*}
B_{11} &=& \xi^2 \kappa^2 (\eta_{11} - \eta_{22}) \\
&+& (k_1^2-k_2^2) (\kb\kb\etab + \kappa \xi \det \etab + \xi^2 \tr \etab), \\
B_{12} &=& \xi^2 \kappa^2 (\eta_{12} + \eta_{21}) + 2 k_1k_2 (\kb\kb\etab + \kappa \xi \det \etab + \xi^2 \tr \etab) .
\end{eqnarray*}
In fact, the expression for energy \eqref{eq:EcpAngles} is another representation of the Eq. \eqref{eq:cas} with explicit indication of the orientation of the molecule. For isotropic molecules with $\alpha_{\mu\nu}  = \alpha \delta_{\mu\nu}$ the energy correction $\Delta \mathcal{E}_\CP = 0$, as should be the case. 

Let us consider general form of conductivity tensor \eqref{eq:etaGen}. We use polar coordinates for $(k_1,k_2)$ and after integrating over angular variable we obtain that $\mathcal{E}_\CP^0 = \mathcal{E}_\CP^s $, where $\mathcal{E}_\CP^s$ is given by Eq. \eqref{eq:Es}, and 
\begin{equation}
	\Delta \mathcal{E}_\CP = - \sin^2\theta (\mathcal{G}_{12}\cos^2\gamma  +\mathcal{G}_{23}),\label{eq:E0-1}
\end{equation}
where 
\begin{eqnarray}
\mathcal{G}_{ij} &=& \int\!\!\!\!\!\int_0^\infty\!\! \frac{k dk d\xi}{4\pi \kappa}  e^{-2 \kappa  a} \frac{\alpha_{ii}-\alpha_{jj}}{W } \left\{ \kappa ^2 (X+Y) \left(\kappa ^2-2 \xi ^2\right)  \right.\nonumber \\
&+& \left. \kappa  \xi  \left(\kappa ^2-3 \xi ^2\right) \left(X^2+X Y+Z^2\right) - X \xi ^4 \right\}. \label{eq:G}
\end{eqnarray}
The energy has no dependence on the angle $\varphi$ due to isotropy of the conductivity plane. The angle $\gamma$ gives contribution in the case $\alpha_{11} \not = \alpha_{22}$, only.  
 
In general, the CP energy depends on three angles $(\theta,\varphi,\gamma)$ which define orientation of the molecule in the space. Therefore, we may define the Casimir torque in relation to each angle \cite{Parsegian:1972:Dielectric,Barash:1978:movdwfbab}
\begin{eqnarray}
M_\theta &=& - \partial_\theta \mathcal{E}_\CP = \sin 2\theta (\mathcal{G}_{12}\cos^2\gamma  +\mathcal{G}_{23}),\nonumber\\
M_\varphi &=& - \partial_\varphi \mathcal{E}_\CP = 0, \nonumber \\
M_\gamma &=& - \partial_\gamma \mathcal{E}_\CP = \sin^2\theta \sin 2\gamma \mathcal{G}_{12}.  \label{eq:Mtorq}
\end{eqnarray}
Zero torque with minima energy $\mathcal{E}_\CP$ describes the equilibrium states.

\subsection{The molecule CO$_2$}

\begin{figure}[htb]
	\includegraphics[width=8 cm ]{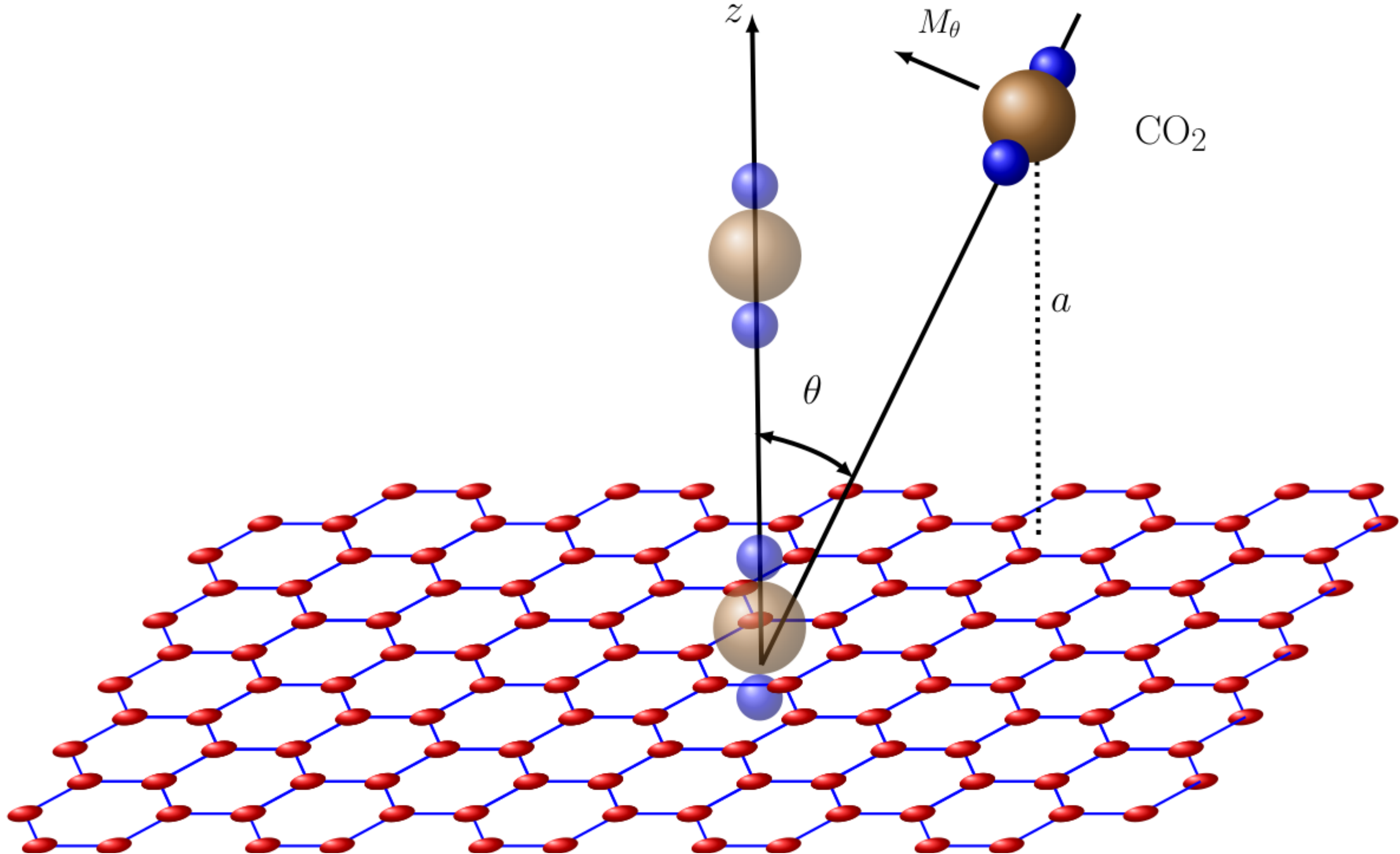}
	\caption{Schematic figure showing the arrangement of the molecule CO$_2$ and the plane. The Casimir torque $M_\theta$ tends to rotate molecule to direction axis $z$ and as the result molecules can go through the membrane.} \label{fig:co2}
\end{figure}

\begin{figure}[htb]
	\includegraphics[width=8 cm ]{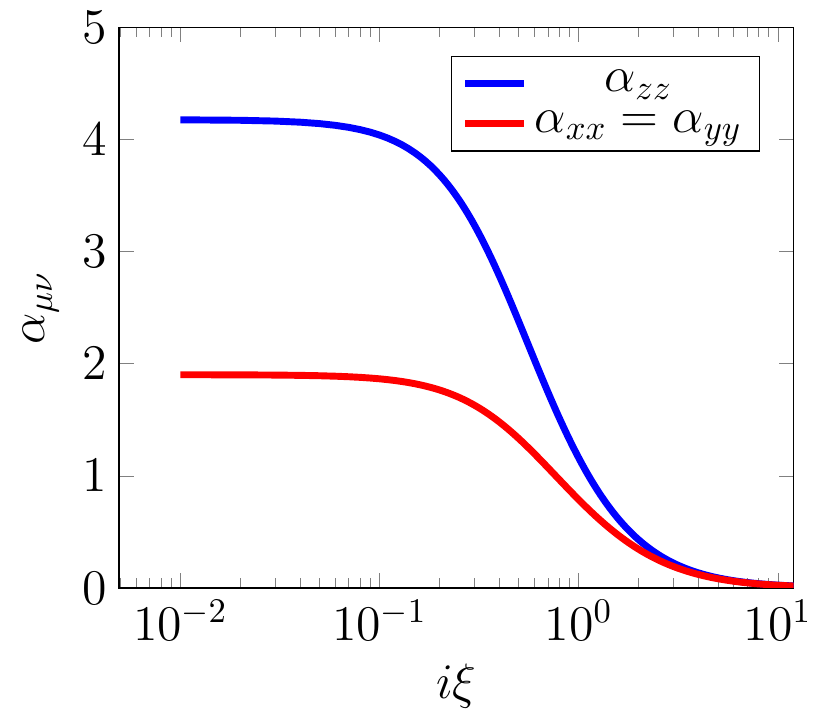}
	\caption{The polarizability of CO$_2$. The numerical data provided by the authors of Ref. \cite{Thiyan:2015:acvdWCPeCO2CH4mnstf}. The imaginary frequency $\xi$ is measured in a.u. and polarizability is measured in \AA$^3$. Here 1 a.u. = $4.13413\cdot 10^{16}\mathrm{rad/s} = 27.212\;\mathrm{eV}$.} \label{fig:alpha}
\end{figure}

The molecule CO$_2$ (see Fig.\,\ref{fig:co2}) has anisotropic polarizability and $\alpha_{11} = \alpha_{22} \not = \alpha_{33}$ (see Fig.\,\ref{fig:alpha}), and for the whole interval of the imaginary frequencies the inequality holds, $\alpha_{33} > \alpha_{22}$. The CP energy depends on $\theta$ only
\begin{equation}
\mathcal{E}_\CP = \mathcal{E}_\CP^0 	- \sin^2\theta \mathcal{G}_{23},
\end{equation}
where $\mathcal{E}_\CP^0 = \mathcal{E}_\CP^s$ and $\mathcal{G}_{23}$ are given by Eqs. \eqref{eq:Es} and \eqref{eq:G}, respectively. The $\sin^2\theta$ dependence has already observed for a dielectric slab in Ref. \cite{Thiyan:2015:acvdWCPeCO2CH4mnstf}.

In this case only $M_\theta$ remains non-zero, $M_\theta = \sin 2\theta\ G_{23}$. The $2\theta$ dependence was also observed in Ref. \cite{Parsegian:1972:Dielectric,Barash:1978:movdwfbab}.  Zero torque, $M=0$, which corresponds to extrema of the energy as function of the angle is realized at the angles $\theta_e = \left(0,\frac{\pi}{2},\pi\right)$ with energies  $\mathcal{E}_\CP = (\mathcal{E}_\CP^0 , \mathcal{E}_\CP^0  - \mathcal{G}_{23}, \mathcal{E}_\CP^0)$. The angles $\theta = 0,\pi$ correspond to perpendicular to the surface position and $\theta = \pi/2$ -- parallel to the surface. Let us consider some specific situations. 

\subsubsection{Constant Hall conductivity}

For pure Hall constant conductivity we set $X=Y=0$ and we obtain from \eqref{eq:EcpAngles}
\begin{eqnarray}
\mathcal{E}_\CP^0 &=& - \frac{Z^2}{1+Z^2} \int_0^\infty \frac{dy y^3}{2\pi a^4} \int_0^1 dx e^{-2y} \nonumber \\
&\times&((1+x^2)\alpha_{22} + (1-x^2) \alpha_{33}),\nonumber \\
\mathcal{G}_{23} &=& \frac{Z^2}{1+Z^2} \int_0^\infty \frac{dy y^3}{4\pi a^4} \int_0^1 dx e^{-2y}\nonumber \\
&\times& (1-3x^2)(\alpha_{22}- \alpha_{33}),
\end{eqnarray}
where $\alpha_{ij} = \alpha_{ij} \left(\frac{\ii x y}{a}\right)$.   

\begin{figure}[htb]
		\includegraphics[width=4.5 cm ]{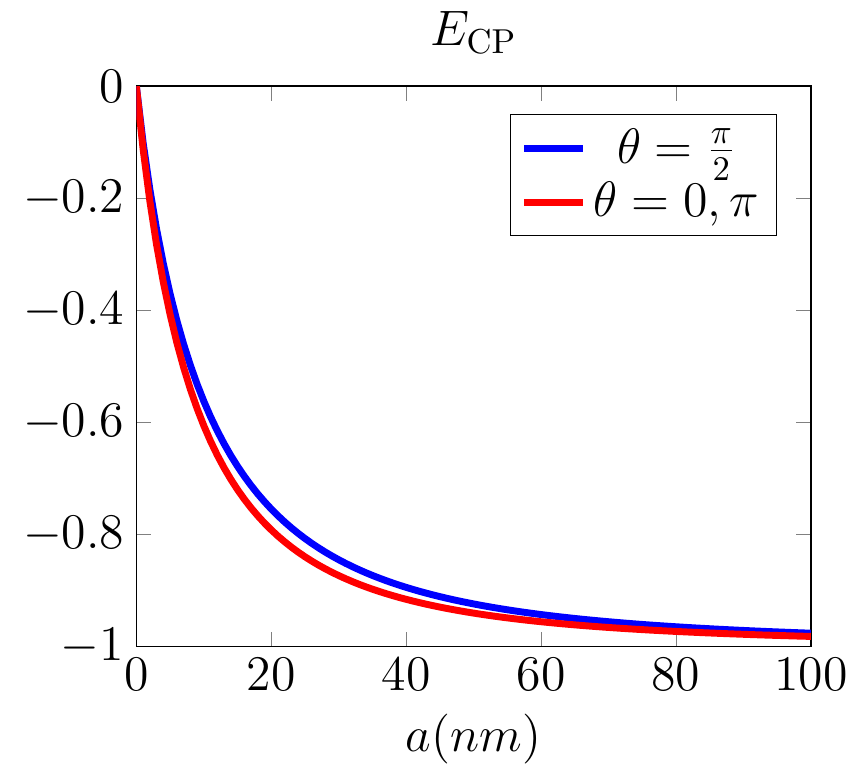}\!\!\!\!\includegraphics[width=4.3 cm ]{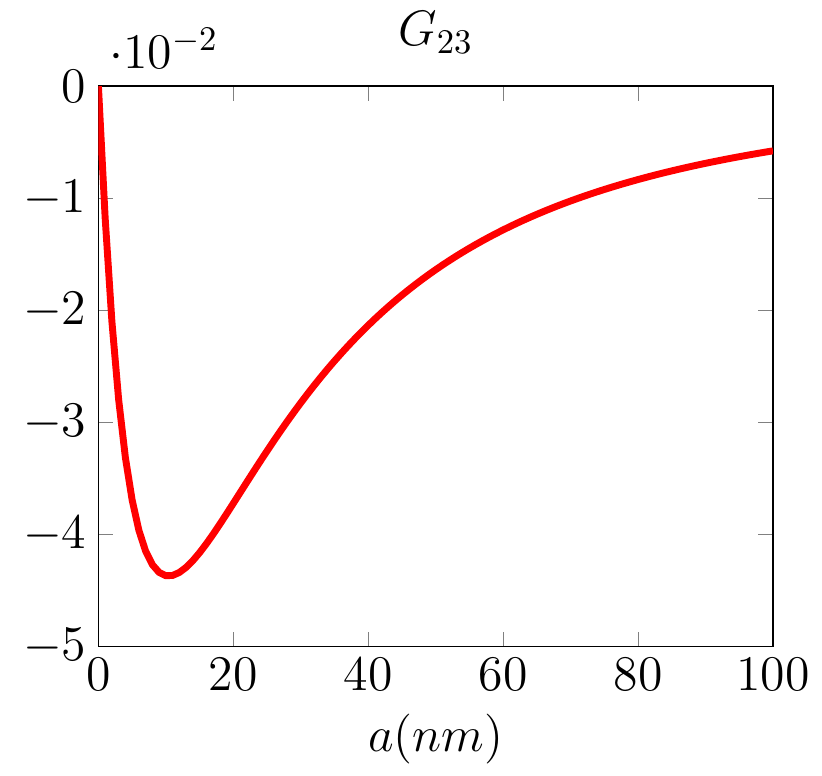}
	\caption{Left panel: the energy $E_\CP$  for extremal angles. The minima energy are realized for $\theta = 0,\pi$.  Right panel: $G_{23}$ contribution, responsible for the angle dependence. It is always negative.} \label{fig:hall}
\end{figure}

For $a\to\infty$ we obtain  (see \eqref{eq:Eid})
\begin{eqnarray}
\mathcal{E}_\CP^0 &\to& \mathcal{E}_\CP^{\infty} =  -\frac{\alpha^\mu_\mu (0)}{8\pi a^4},\nonumber \\
\mathcal{G}_{23} &=& {\cal O}(a^{-5}).
\end{eqnarray}
For numerical evaluation we normalize the energy on its asymptotic value $|\mathcal{E}_\CP^{\infty}| = \frac{\alpha^\mu_\mu(0)}{8\pi a^4}$: $\mathcal{E}_\CP/|\mathcal{E}_\CP^{\infty}| = \frac{Z^2}{1+Z^2}  E_\CP$ and $\mathcal{G}_{23}/|\mathcal{E}_\CP^{\infty}| = \frac{Z^2}{1+Z^2}  G_{23}$. The plots of the whole energy $E_\CP$ and separately $G_{23}$ contribution, responsible for the $\theta$ dependence, are shown in Fig.\,\ref{fig:hall}. 
We observe that the CP energy with angles $\theta = 0,\pi$ is smaller than the energy for $\theta = \pi/2$  for any distance between atom and plane.  It means that equilibrium position of molecules are perpendicular to the plane.

\subsubsection{Graphene}

In this case $X=\eta_\te, Y= \eta_\tm   - \eta_\te$ and $Z=0$, where $\eta_\te$  and $\eta_\tm  $ are conductivities TE and TM modes, respectively \cite{Khusnutdinov:2019:lteotcfefaaiwacp}. The energy for zero temperature reads 
\begin{eqnarray}
\mathcal{E}_\CP^0 &=& -\int_0^\infty \frac{kdk}{2\pi}  \int_0^\infty \frac{d\xi}{\kappa} e^{-2 \kappa  a} \nonumber \allowdisplaybreaks \\
&\times&\left(-\alpha_{22} \xi^2 r_\te + (\alpha_{22} \kappa^2 + \alpha_{33} (\kappa^2 - \xi^2)) r_\tm  \right),\nonumber \allowdisplaybreaks \\
\mathcal{G}_{23} &=& \int_0^\infty \frac{k dk}{4\pi}  \int_0^\infty \frac{d\xi}{\kappa} e^{-2 \kappa  a} (\alpha_{22}-\alpha_{33})\nonumber \allowdisplaybreaks \\
&\times& \left(\xi ^2 r_\te + (\kappa ^2 - 2 \xi ^2) r_\tm  \right), \label{eq:Egra}
\end{eqnarray} 
where 
\begin{equation}
r_\te =  - \frac{\eta_\te}{\eta_\te + \frac{\kappa}{\xi}}, r_\tm   =  \frac{\eta_\tm  }{\eta_\tm   + \frac{\xi}{\kappa}},
\end{equation}
are reflection coefficients for graphene and $\eta_{\te,\tm} = 2\pi \sigma_{\te,\tm}$. At zero temperature, the energy can also depend, in general, on the chemical potential $\mu$  \cite{Bordag:2009:CibapcagdbtDm}. In the gapeless case the conductivity of TM and TE modes have the following form \cite{Emelianova:2020:Cedapsc}
\begin{eqnarray}
	\frac{\eta_\tm}{\eta_\gr} &=& \frac{\xi}{k_F} + \frac{8}{\pi k_F} \re \int_0^{|\mu|} dz \frac{qr -\xi k_F}{k_F r + q \xi} , \nonumber \allowdisplaybreaks \\
	\frac{\eta_\te}{\eta_\gr} &=& \frac{k_F}{\xi} + \frac{8}{\pi \xi} \re \int_0^{|\mu|} dz \frac{q(q^2 -\xi^2)}{r(q k_F  + \xi r)} , \label{eq:etas}
\end{eqnarray}
where $k_F = \sqrt{\xi^2 + v_F^2 k^2}$, $q= \xi - \ii z$, $r= \sqrt{q^2 + v_F^2  k^2}$, and $\eta_\gr = 2\pi e^2/4$ is the universal conductivity of graphene.  

The non-zero temperature maybe included by changing $\int_0^\infty d\xi \to 2\pi T \mathop{\sum\nolimits'}_{n=0}^\infty$ and $\xi \to \xi_n$, where $\xi_n = 2\pi n T$ are the Matsubara frequencies:
\begin{eqnarray}
	\mathcal{E}_\CP^0 &=& -T \mathop{\sum\nolimits'}_{n=0}^\infty\int_0^\infty \frac{kdk}{\kappa_n} e^{-2 \kappa_n  a} \nonumber \allowdisplaybreaks \\
	&\times&\left(-\alpha_{22}^n \xi_n^2 r_\te^n + (\alpha_{22}^n \kappa_n^2 + \alpha_{33}^n (\kappa_n^2 - \xi_n^2)) r_\tm^n  \right),\nonumber \allowdisplaybreaks \\
	\mathcal{G}_{23} &=& T \mathop{\sum\nolimits'}_{n=0}^\infty\int_0^\infty \frac{kdk}{2\kappa_n}  e^{-2 \kappa_n  a} (\alpha_{22}^n-\alpha_{33}^n)\nonumber \allowdisplaybreaks \\
	&\times& \left(\xi_n^2 r_\te^n + (\kappa_n^2 - 2 \xi_n^2) r_\tm^n  \right), \label{eq:Egra-1}
\end{eqnarray} 
where $\kappa_n = \sqrt{k^2+\xi_n^2}$ and $\alpha_{\mu\nu}^n = \alpha_{\mu\nu}(\ii \xi_n)$. The conductivity of the graphene in this case read 
\begin{eqnarray}
	\frac{\eta_\tm}{\eta_\gr} &=& \frac{\xi}{k_F} + \frac{8}{\pi k_F} \re \int_0^\infty dz \frac{qr -\xi k_F}{k_F r + q \xi} \Theta(z), \nonumber \\
	\frac{\eta_\te}{\eta_\gr} &=& \frac{k_F}{\xi} + \frac{8}{\pi \xi} \re \int_0^\infty dz \frac{q(q^2 -\xi^2)}{r(q k_F  + \xi r)} \Theta(z), \label{eq:etamu0} 
\end{eqnarray} 
where 
\begin{equation}
	\Theta(z) = \frac{1}{e^{\frac{z+\mu}{T}} + 1} + \frac{1}{e^{\frac{z-\mu}{T}} + 1}.
\end{equation}

Let us consider the asymptotic $a\to \infty$. In the case of zero temperature we change integrands $k \to k/a$, $\xi \to \xi/a$ in \eqref{eq:Egra} and take formal limit $a\to \infty$. In this limit we obtain 
\begin{eqnarray}
	\frac{\eta_\tm}{\eta_\gr} &\to & \frac{8 a|\mu| \xi (k_F+\xi)}{\pi k^2 k_F v_F^2}  \to \infty, \nonumber \\
	\frac{\eta_\te}{\eta_\gr} &\to& -\frac{8 a|\mu| \xi (k_F+\xi)}{\pi k^2 k_F v_F^2} \to -\infty.\label{eq:etaas}
\end{eqnarray}
The reflection coefficients become those of a perfect metal, $r_\tm = 1$ and $r_\te = -1$ and 
\begin{equation}
	\mathcal{E}_\CP^0 /\mathcal{E}_\CP^\infty \to 1,\ \mathcal{G}_{23}/\mathcal{E}_\CP^\infty \to 0. 
\end{equation}
This regime is realized in the case $a |\mu| \gg 1$. If $\mu =0$, the conductivities are given by the firsts terms in Eq. \eqref{eq:etamu0} and $r_\tm \not = 1$ and $r_\te \not = -1$. Numerically, we obtain 
\begin{eqnarray}
	\mathcal{E}_\CP^0 /\mathcal{E}_\CP^\infty &\to& \frac{\alpha_{22}(0)I_1  + \alpha_{33}(0)I_2}{\alpha^\nu_\nu(0)} \approx 0.0542,\nonumber \\ \mathcal{G}_{23}/\mathcal{E}_\CP^\infty &\to& - \frac{\alpha_{22}(0) - \alpha_{33}(0)}{\alpha^\nu_\nu(0)} I_3 \approx 0.0077,
\end{eqnarray}
where $I_1 \approx 0.079, I_2 \approx 0.067$ and $I_3 \approx 0.027$.

In the case of non-zero temperature, the main contribution comes from zero term of sum in \eqref{eq:Egra-1}:
\begin{eqnarray}
	\mathcal{E}_\CP^0/\mathcal{E}_\CP^\infty &\to& \pi a T\frac{\alpha_{22}(0) + \alpha_{33}(0)}{\alpha^\nu_\nu(0)} \approx \frac{ 2\pi a T}{3} 1.14, \nonumber \\
	\mathcal{G}_{23}/\mathcal{E}_\CP^\infty &\to& -\pi a T\frac{\alpha_{22}(0) - \alpha_{33}(0)}{2\alpha^\nu_\nu(0)} \approx \frac{ 2\pi a T}{3} 0.21. \label{eq:Egra-2}
\end{eqnarray} 

\begin{figure}[htb]
	\includegraphics[width=4.3 cm ]{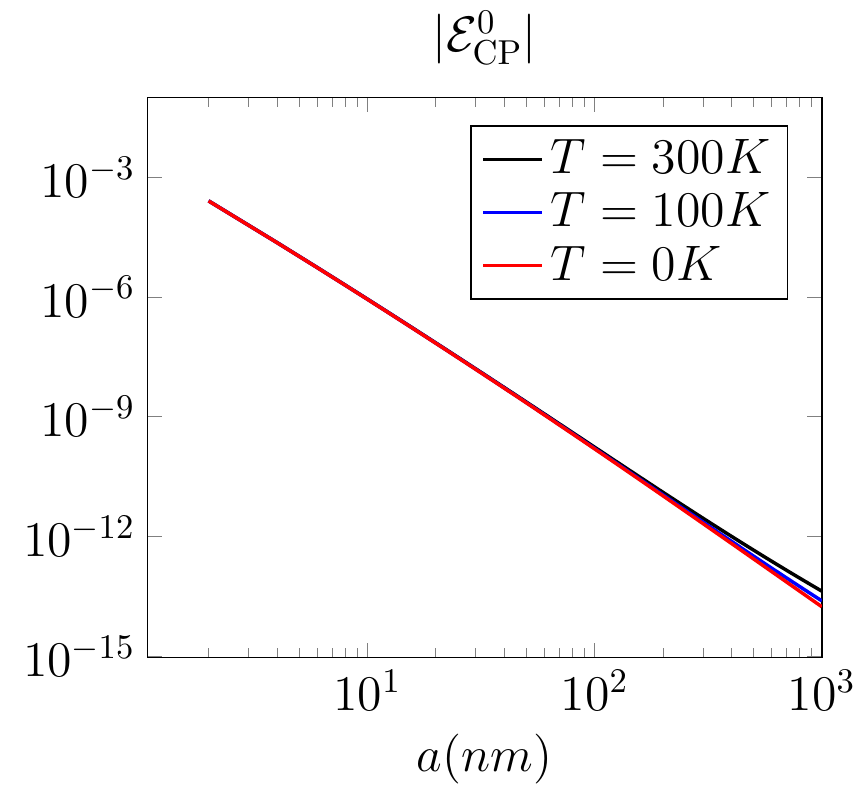}\includegraphics[width=4.3 cm ]{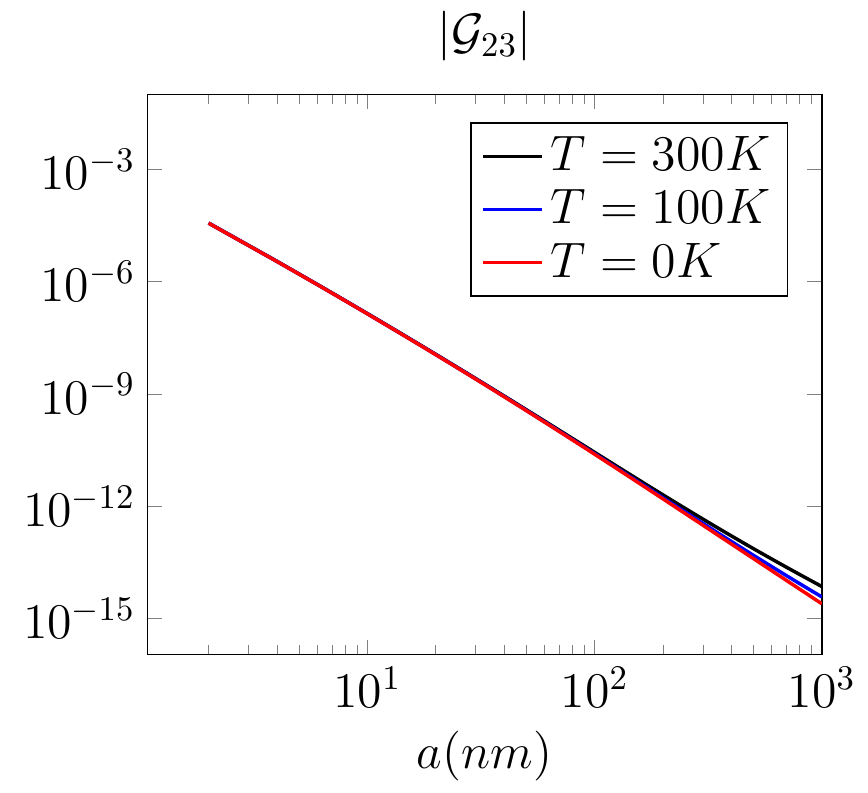}
	\caption{Left panel: the energy $|\mathcal{E}_\CP^0|$  in double log scale  for  different temperatures. Right panel: the energy $|\mathcal{G}_\CP|$  in double log format  for  various temperatures. For small distances the energies $\sim 1/a^4$. For great distance and non-zero temperatures the energies $\sim T/a^3$.} \label{fig:graph1}
\end{figure}
The numerical analysis reveals that the function $\mathcal{G}_{23}$ is negative for any distance between plane and molecule. It means that the minima energy are realized for $\theta = 0,\pi$. The module of energies are show in Fig.\,\ref{fig:graph1} in double log scale in range $a \in [1,500]nm$ and temperatures $T=0,100,300K$. We observe that for zero temperature the energies $\mathcal{E}_\CP^0$ and $\mathcal{G}_{23}$ proportional to $a^{-4}$. The non-zero temperature changes behaviour at large distance between graphene and molecule to $T/a^3$ according to \eqref{eq:Egra-2}.  
  
\begin{figure}
	\includegraphics[width=2.9 cm ]{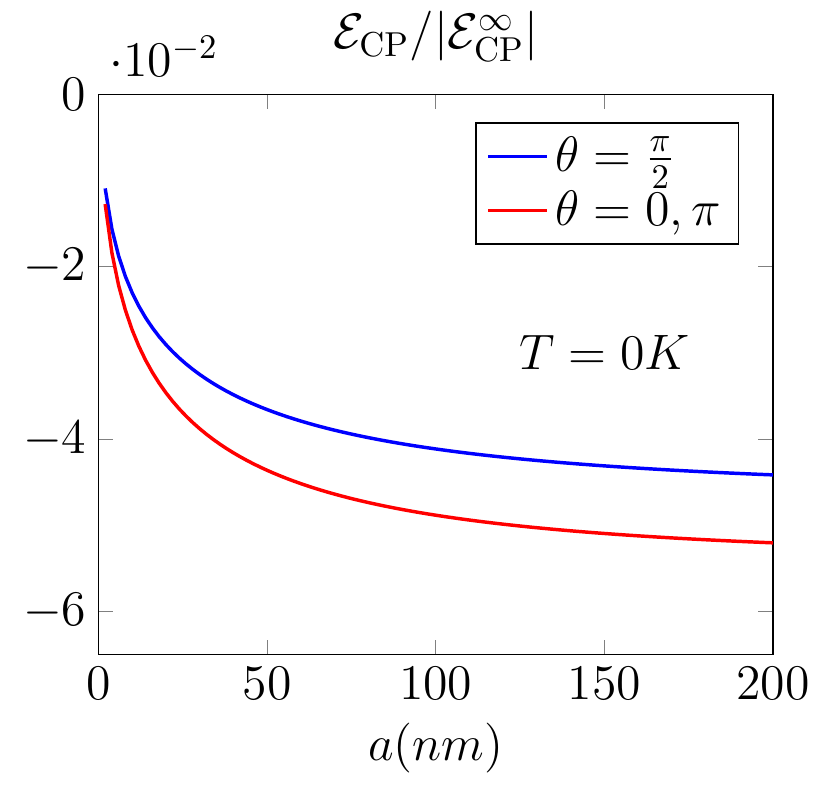}\!\includegraphics[width=2.9 cm ]{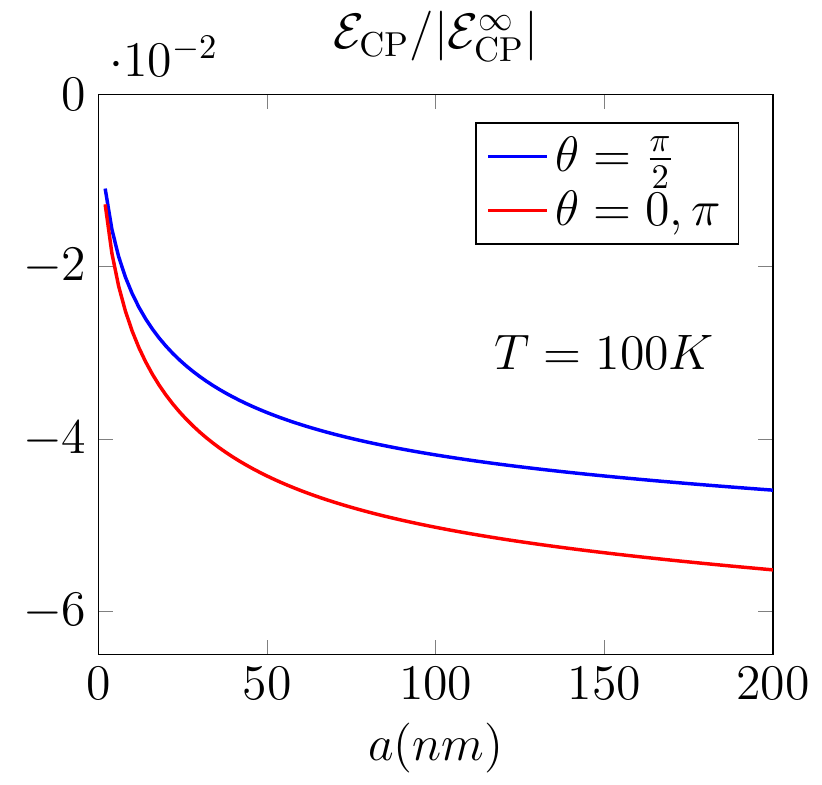}\!\includegraphics[width=2.9 cm ]{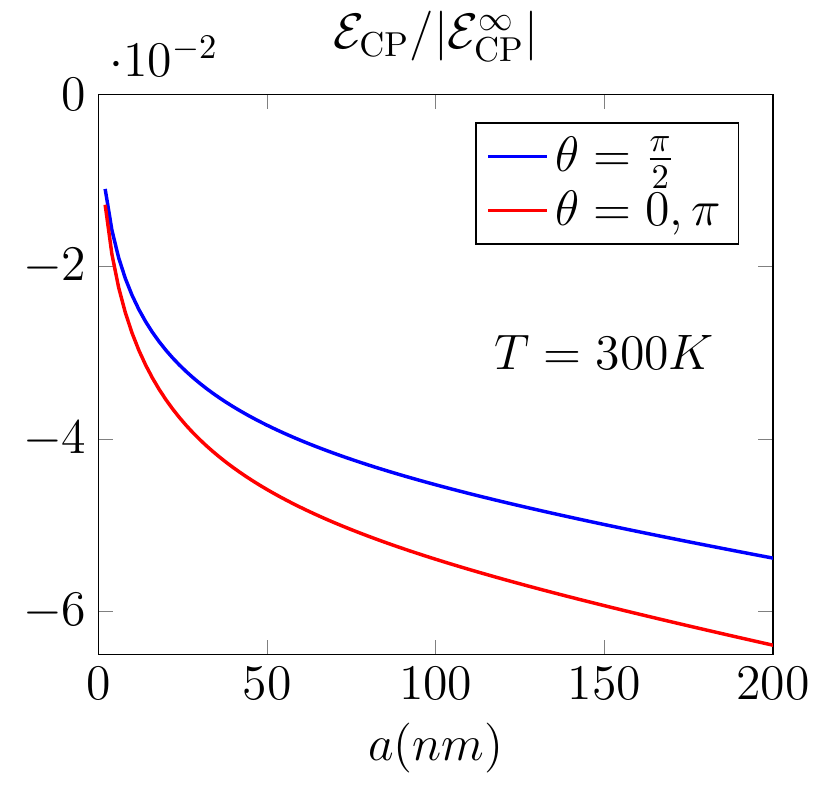}
	\caption{The energy $\mathcal{E}_\CP/|\mathcal{E}^\infty_\CP|$ for temperature $T=0K$ (left panel), $T=100K$ (middle panel) and $T=300K$ (right panel). The minima of the energy are realized for $\theta = 0,\pi$.} \label{fig:graph2}
\end{figure}

The energy for extremal angles $\theta = 0, \pi/2,\pi$ and for temperatures $T=0,100,300K$ are shown in Fig.\,\ref{fig:graph2}. In all situations the minima of energy are realized for $\theta = 0,\pi$, when the molecule is perpendicular to the surface (see Fig.\,\ref{fig:co2}).  

\section{Summary and conclusion} \label{Sec:Summary}

We considered the Casimir-Polder energy for an anisotropic molecule near a conductive plane with tensorial surface conductivity.  The CP energy is a sum of the contributions from symmetric and antisymmetric \eqref{eq:Asym} parts of polarizability.  The latter is zero either for symmetric polarizability tensor or for symmetric conductivity tensor. The CP energy is represented in the form with manifest dependence on the orientation of the molecule \eqref{eq:EcpAngles}. The Casimir torques is given by derivatives with respect to angles of orientation of the molecule \eqref{eq:Mtorq}. The energy has no dependence on the spherical angle $\varphi$ due to the freedom of choosing the origin of coordinates and dependence on $\theta$ is of the type $\sin^2\theta$. The dependence on the angle $\gamma$ is only present in the case $\alpha_{11}\not = \alpha_{22}$. This angle characterizes a rotation around radius vector to molecule (see Fig.\,\ref{fig:sis}). 

The molecule CO$_2$ has tensor polarizability with $\alpha_{11} = \alpha_{22} \not = \alpha_{33}$ and $\alpha_{33} > \alpha_{11}$ for any frequency. For this reason the energy depends on $\theta$ as $\sin^2\theta$ and the minima of the energy and zero torque are realized for $\theta = 0,\pi$, when the molecule is perpendicular to surface and maximum is for $\theta = \pi/2$. The same result has been obtained for molecule CO$_2$ and dielectric slab in Ref. \cite{Thiyan:2015:acvdWCPeCO2CH4mnstf}. Numerical results for graphene and molecule CO$_2$ are shown in Fig.\,\ref{fig:graph2} for different temperatures.  

\begin{figure}[h]
	\includegraphics[width=4.25 cm ]{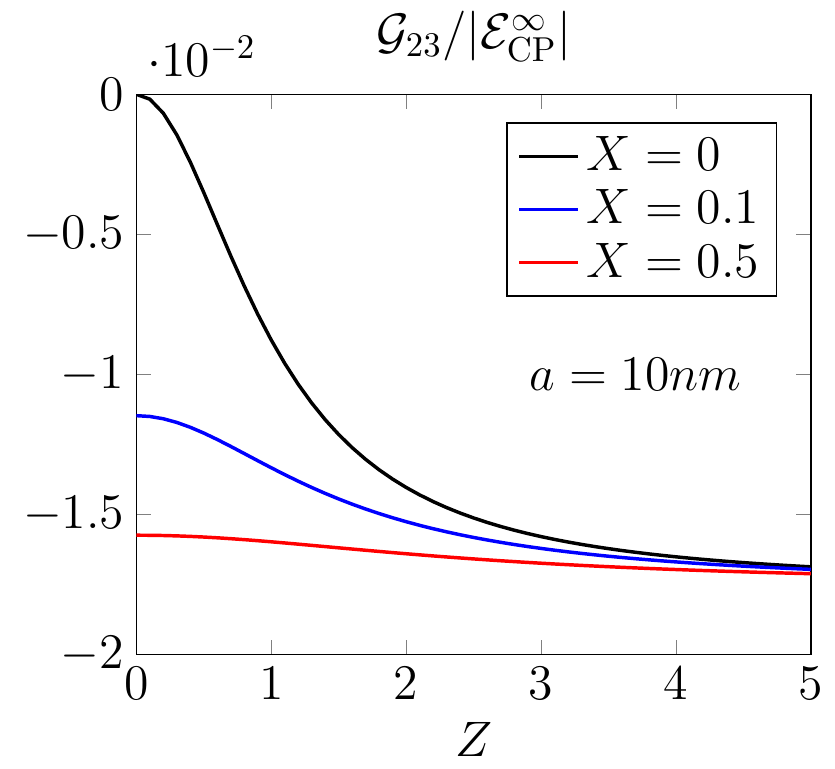}\includegraphics[width=4.4 cm ]{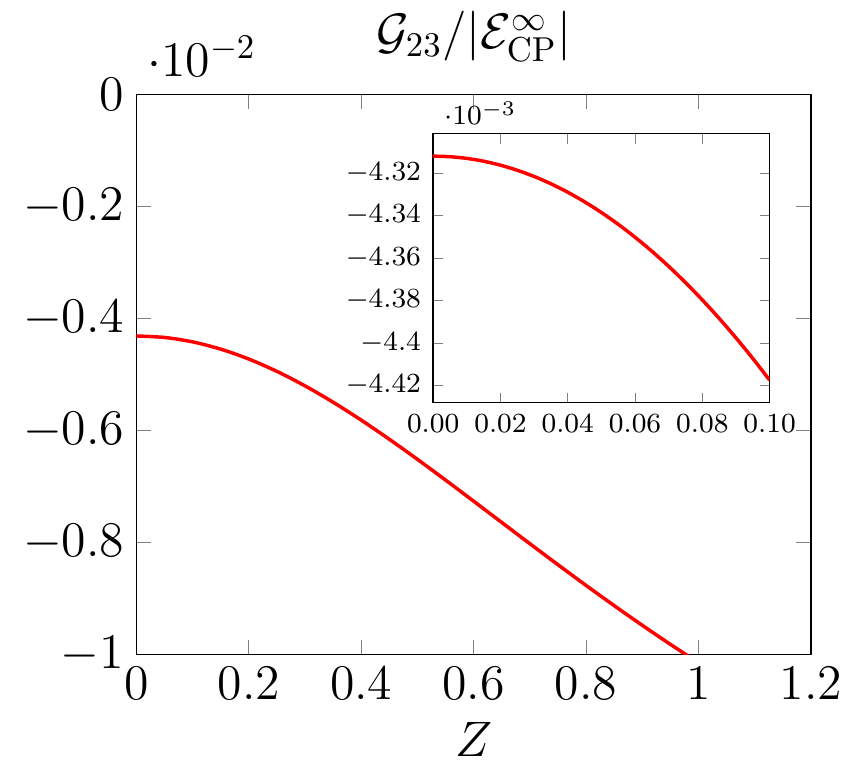}
	\caption{The dependence of the torque amplitude ${\cal G}_{23}$ on the anti-symmetric (Hall) part, $Z$, of the conductivity in the case of constant symmetric conductivity (left panel) and for graphene (right panel).} \label{fig:z-dep}
\end{figure}

Separating CO$_2$ from the atmosphere is a pressing technological issue, in many cases addressed with usage of graphene and graphene oxide membranes \cite{ALI201983,YOO201739}, including those of single-layered graphene \cite{Zhou2018}. Our study revealed that the Casimir torque acting on the molecules tends to shift them into a position favoring easier penetration through such membranes -- with the smallest cross-section side facing the interface. In the studied cases, see Fig.\,\ref{fig:z-dep}, the torque is showed to be enhanced by enlarging the Hall conductivity contribution of the conductivity tensor of the membrane, which is notably pronounced in the case of graphene. It permits us to conjecture, that creating membranes with larger antisymmetric conductivities might benefit the CO$_2$ separation technologies. 

\begin{acknowledgments}
NK was supported in part by the grants 2019/10719-9, 2016/03319-6 of S\~ao Paulo Research Foundation (FAPESP) and by the Russian Foundation for Basic Research Grant No. 19-02-00496-a. One of us (NK) is grateful to Valery Marachevsky for fruitful discussions. The authors thank to the authors of the Ref. \cite{Thiyan:2015:acvdWCPeCO2CH4mnstf} for numerical data for polarizability tensor of the molecule CO$_2$. 
\end{acknowledgments}

\appendix 

\section{Derivation of CP energy}\label{Sec:Ap1}
\textbf{I. Eigenproblem}

The Maxwell equations in anisotropic media  give a dispersion relation which has, in general, $4$  distinctive roots, see below in \eqref{eq:roots}, $k_3 = k_3 (k_1,k_2,\omega)$ and $4$ corresponding distinct amplitudes $\bm{\mathcal{E}}_{n}$ and $\bm{\mathcal{H}}_{n}$ ($n=1,2,3,4$). To obtain this dispersion relation we search for solutions of the Maxwell equations in the plane waves form
\begin{equation}
\mathbf{E} = e^{\ii k_1 x + \ii k_2 y + \ii k_3 z -\ii \omega t} \bm{\mathcal{E}}, \mathbf{H} = e^{\ii k_1 x + \ii k_2 y + \ii k_3 z -\ii \omega t} \bm{\mathcal{H}},\label{eq:Ep}
\end{equation}
with constant amplitudes $\bm{\mathcal{E}}$ and $\bm{\mathcal{H}}$. 

The equations can be represented in the form of an eigenproblem 
\begin{equation}\label{eq:eigen}
\mathbf{M}\cdot \bm{v} = k_3 \bm{v},
\end{equation}
where the matrix $\mathbf{M}$ is given by 
\begin{equation}\label{eq:M}
\mathbf{M} = \scriptstyle
\begin{bmatrix}
-k_1 \frac{\varepsilon_{31}}{\varepsilon_{33}} & -k_1 \frac{\varepsilon_{32}}{\varepsilon_{33}} & \frac{k_1 k_2}{\omega \varepsilon_{33}} & \omega - \frac{k_1^2}{\omega \varepsilon_{33}} \\
-k_2 \frac{\varepsilon_{31}}{\varepsilon_{33}} & -k_2 \frac{\varepsilon_{32}}{\varepsilon_{33}} & -\omega + \frac{k_2^2}{\omega \varepsilon_{33}}  & -\frac{k_1 k_2}{\omega \varepsilon_{33}}\\
-\frac{k_1 k_2}{\omega} - \frac{\omega e_{12}}{\varepsilon_{33}} & \frac{k_1^2}{\omega} - \frac{\omega e_{11}}{\varepsilon_{33}} & -k_2 \frac{\varepsilon_{23}}{\varepsilon_{33}} & k_1 \frac{\varepsilon_{23}}{\varepsilon_{33}}\\
-\frac{k_2^2}{\omega} + \frac{\omega e_{22}}{\varepsilon_{33}} & \frac{k_1 k_2}{\omega} + \frac{\omega e_{21}}{\varepsilon_{33}} & k_2 \frac{\varepsilon_{13}}{\varepsilon_{33}} & -k_1 \frac{\varepsilon_{13}}{\varepsilon_{33}}
\end{bmatrix},
\end{equation}
and $4$-component "vector" $\bm{v}$ reads
\begin{equation}
\bm{v} = 
\begin{pmatrix}
\mathcal{E}_x\\
\mathcal{E}_y\\
\mathcal{H}_x\\
\mathcal{H}_y
\end{pmatrix} \equiv 
\begin{pmatrix}
\eb \\
\hb
\end{pmatrix},
\label{e,h}
\end{equation}
which contains the tangent components of electromagnetic field.  Here, $e_{\mu\nu}$ is a matrix of minors of the elements of $\varepsilon_{\mu\nu}$. To find a solution we need for these components of the electromagnetic field, only, since the third components are readily obtained as
\begin{eqnarray}
\mathcal{E}_z &=& \frac{1}{\omega \varepsilon_{33}} \left(k_2 \mathcal{H}_x - k_2 \mathcal{H}_y - \omega \left(\varepsilon_{31} \mathcal{E}_x + \varepsilon_{32} \mathcal{E}_y\right)\right),\nonumber \\
\mathcal{H}_z &=& \frac{1}{\omega} \left(k_1 \mathcal{E}_y - k_2 \mathcal{E}_x\right).
\end{eqnarray}

The dispersion relation, $k_3 = k_3 ( \kb ,\omega)$, is solution of the 4-th degree equation
\begin{equation}\label{eq:roots}
\varepsilon_{33}\det (\mathbf{M}- k_3 \mathbf{I}) 
	\equiv 
		\varepsilon_{33} k_3^4 + a_1 k_3^3 + a_2 k_3^2 + a_3 k_3 + a_4 = 0,
\end{equation}
where coefficients
\begin{eqnarray*}
a_1 &=&  k^i (\varepsilon_{i3} + \varepsilon_{3i}), \nonumber \allowdisplaybreaks \\
a_2 &=& k^2 \varepsilon_{33} +   \kb  \kb \bm{\varepsilon} - \omega^2 (e_{11}+e_{22}), \nonumber \allowdisplaybreaks \\
a_3 &=&  \kb ^2 k^i (\varepsilon_{i3} + \varepsilon_{3i})  + \omega^2  k^i (\varepsilon^{-1}_{i3} + \varepsilon^{-1}_{3i}) \det\bm{\varepsilon},\nonumber \allowdisplaybreaks\\
a_4 &=&  \kb ^2  \kb  \kb \bm{\varepsilon} - \omega^2 ( \kb ^2 e_{33} + \varepsilon_{33}  \kb  \kb \bm{\varepsilon} -  k^i \varepsilon_{i3} k^j \varepsilon_{3j})\allowdisplaybreaks\\
&+& \omega^4 \det\bm{\varepsilon},
\end{eqnarray*}
are invariants over rotation in plane $z=0$.  We denote the roots of \eqref{eq:roots} as $\kappa_n$. We choose numeration of roots such that in the case of vacuum $\kappa_{1,2}\to + k_z$ and $\kappa_{3,4} \to -k_z$, where $k_z = \sqrt{\omega^2 -  \kb ^2}$.  

In the case of vacuum, $\varepsilon_{\mu\nu} = \delta_{\mu\nu}$, the dispersion relation has double-degenerate roots  $\kappa_{1,2} = +k_z$ and $\kappa_{3,4} = -k_z$. Corresponding eigenvectors read 
\begin{equation}
\vb_1^r = 
\begin{pmatrix}
1 \\
0\\
-\frac{k_1 k_2}{\omega k_z} \\
\frac{k_1^2 + k_z^2}{\omega k_z}
\end{pmatrix},\ 
\vb_2^r = 
\begin{pmatrix}
0 \\
1\\
-\frac{k_2^2 + k_z^2}{\omega k_z} \\
\frac{k_1 k_2}{\omega k_z}
\end{pmatrix},
\end{equation}
and $(\bm{v}_3^r, \bm{v}_4^r) = (\bm{v}_3^r, \bm{v}_4^r)_{k_z \to - k_z}$.

Then, the general form of field is a linear combination of these solutions
\begin{equation}\label{eq:vv}
\vb^r = e^{\ii k_z z} v_1^r \vb_1^r + e^{\ii k_z z} v_2^r \vb_2^r  + e^{-\ii k_z z} v_3^r \vb_3^r + e^{-\ii k_z z} v_4^r \vb_4^r,
\end{equation}
with common factor $e^{\ii k_1 x + \ii k_2 y  -\ii \omega t}$. 

In the non-vacuum case the amplitudes read  
\begin{widetext}
\begin{eqnarray}
\bm{v}_1 &=& 
\begin{pmatrix}
1 \\[1em]
\frac{k_2 (k_1 \xi_3 - k_3 \xi_1) + \omega^2 (k_3 e_{12} - k_1 e_{32})}{k_3 (kk\varepsilon) + k_1 (k_1 \xi_3 - k_3 \xi_1) + \omega^2 (k_1 e_{31} - k_3 e_{11})}\\[1em]
\frac{- k_1 k_2 (kk\varepsilon) + \omega^2 (k_1 k_3 e_{32} + k_1 k_2 e_{33} - k_3^2 e_{12} - k_2 k_3 e_{13})}{\omega(k_3 (kk\varepsilon) + k_1 (k_1 \xi_3 - k_3 \xi_1) + \omega^2 (k_1 e_{31} - k_3 e_{11}))} \\[1em]
\frac{-k_2^2 (kk\varepsilon) + \omega^2 (k_1 k_2 (e_{12} + e_{21}) + (k_1^2 + k_3^2) e_{22} + k_2 k_3 (e_{32} + e_{23}) + k_2^2 (e_{11} + e_{33})) - \omega^4 \det\bm{\varepsilon}}{\omega(k_3 (kk\varepsilon) + k_1 (k_1 \xi_3 - k_3 \xi_1) + \omega^2 (k_1 e_{31} - k_3 e_{11}))}
\end{pmatrix},\ k_3 \to \kappa_1,\nonumber \\ [1em] 
\bm{v}_2 &=& 
\begin{pmatrix}
\frac{k_1 (k_2 \xi_3 - k_3 \xi_2) + \omega^2 (k_3 e_{21} + k_2 e_{31})}{k_3 (kk\varepsilon) + k_2 (k_2 \xi_3 - k_3 \xi_2) - \omega^2 (k_2 e_{32} + k_3 e_{22})} \\[1em]
1\\[1em]
\frac{k_1^2 (kk\varepsilon) + \omega^2 (-k_1 k_2 (e_{12} + e_{21}) - (k_2^2 + k_3^2) e_{11} + k_1 k_3 (e_{31} + e_{13}) - k_1^2 (e_{22} + e_{33})) + \omega^4 \det \bm{\varepsilon}}{\omega(k_3 (kk\varepsilon) + k_2 (k_2 \xi_3 - k_3 \xi_2) - \omega^2 (k_2 e_{32} + k_3 e_{22}))} \\[1em]
\frac{k_1 k_2 (kk\varepsilon) + \omega^2 (-k_1 k_3 e_{23} - k_1 k_2 e_{33} + k_3^2 e_{21} + k_2 k_3 e_{31})}{\omega(k_3 (kk\varepsilon) + k_2 (k_2 \xi_3 - k_3 \xi_2) - \omega^2 (k_2 e_{32} + k_3 e_{22}))} \\[1em]
\end{pmatrix},\ k_3 \to \kappa_2, \label{eq:eigenvectors}
\end{eqnarray}
\end{widetext}
where $e_{\mu\nu}$, as before, is a matrix of minors of $\bm{\varepsilon}_{\mu\nu}$, $\xi_\nu = k^\mu \varepsilon_{\mu\nu}$ and $(kk\varepsilon) = k^\mu k^\nu \varepsilon_{\mu\nu}$. Also $\vb_3 = \left.\vb_1 \right|_{k_3 \to \kappa_3}$, $\vb_4 = \left.\vb_2 \right|_{k_3 \to \kappa_4}$. 

The general solution is a linear combination of these 4 solutions 
\begin{equation}\label{eq:vnv}
\vb = e^{\ii \kappa_1 z} v_1^l \vb_1 + e^{\ii \kappa_2 z} v_2^l \vb_2  + e^{\ii \kappa_3 z} v_3^l \vb_3 + e^{\ii \kappa_ 4 z} v_4^l \vb_4.
\end{equation}

In the vacuum limit $\varepsilon_{\mu\nu} = \delta_{\mu\nu} + \epsilon \alpha_{\mu\nu}$ and $k_3^{(1,2)} = + k_z + \epsilon \delta k_3^{(1,2)}$, $k_3^{(3,4)} = - k_z + \epsilon \delta k_3^{(3,4)}$ we obtain 
\begin{eqnarray}
	\bm{v}_1 &\to& \bm{v}_1^0 + c_1 \bm{v}_2^0,\ \bm{v}_2 \to c_2 \bm{v}_1^0 + \bm{v}_2^0,\nonumber \\
	\bm{v}_3 &\to& \bm{v}_3^0 + c_3 \bm{v}_4^0, \ \bm{v}_4 \to  c_4 \bm{v}_3^0 + \bm{v}_4^0, \label{eq:vacuumlimit}
\end{eqnarray}
where 
\begin{eqnarray}
	c_1 &=& -\frac{\omega^2 (k_z \alpha_{21} - k_1 \alpha_{23}) + k_2 (k_1 \zeta_3 - k_z \zeta_1)}{\omega^2 (k_z \alpha_{22} - k_2 \alpha_{23}) + k_2 (k_2 \zeta_3 - k_z \zeta_2) - 2 k_z^2 \delta k_3^{(1)}},\nonumber\allowdisplaybreaks \\
	c_2 &=& -\frac{\omega^2 (k_z \alpha_{12} - k_2 \alpha_{13}) + k_1 (k_2 \zeta_3 - k_z \zeta_2)}{\omega^2 (k_z \alpha_{11} - k_1 \alpha_{13}) + k_1 (k_1 \zeta_3 - k_z \zeta_1) - 2 k_z^2 \delta k_3^{(2)}},\nonumber\allowdisplaybreaks \\
	c_3 &=& \left. c_1\right|_{k_z \to - k_z, \delta k_3^{(1)} \to \delta k_3^{(3)}}, \nonumber\allowdisplaybreaks \\
	c_4 &=& \left. c_2\right|_{k_z \to - k_z, \delta k_3^{(2)} \to \delta k_3^{(4)}},
\end{eqnarray}
and $\zeta_i = k^n \alpha_{ni}$.  We note that the eigenvectors \eqref{eq:eigenvectors} have chosen in that form that in the vacuum limit these eigenvectors become a linear combination of vacuum vectors in agreement with quantum mechanical perturbation theory in the degenerate case. 

\bigskip
\textbf{II Scattering problem}

Let us consider now a general scattering problem with matter described by a dielectric permittivity  $\varepsilon_{\mu\nu}$ in the left (index $l$) of the boundary $z=0$ and vacuum, $\varepsilon_{\mu\nu} = \delta_{\mu\nu}$, in the right (index $r$). The field has the following structure at the left of the boundary (inside matter): 
\begin{eqnarray*}
	\mathbf{E}_l &=& e^{\ii \kappa_1 z} A_{i}^{l} \bm{\mathcal{E}}_{1}^l + e^{\ii \kappa_2 z} B_{i}^{l} \bm{\mathcal{E}}_{2}^l  + e^{\ii \kappa_3 z} A_{o}^{l} \bm{\mathcal{E}}_{3}^l + e^{\ii \kappa_4 z} B_{o}^{l} \bm{\mathcal{E}}_{4}^l, \nonumber \\
	\mathbf{H}_l &=& e^{\ii \kappa_1 z} A_{i}^{l} \bm{\mathcal{H}}_{1}^l \! + e^{\ii \kappa_2 z} B_{i}^{l} \bm{\mathcal{H}}_{2}^l \! + e^{\ii \kappa_3 z} A_{o}^{l} \bm{\mathcal{H}}_{3}^l \! + e^{\ii \kappa_4 z} B_{o}^{l} \bm{\mathcal{H}}_{4}^l,  
\end{eqnarray*}
and on the right of boundary (in vacuum)
\begin{eqnarray*}
	\mathbf{E}_r &=& e^{\ii k_z z} A_{o}^{r} \bm{\mathcal{E}}_{1}^r + e^{\ii k_z z} B_{o}^{r} \bm{\mathcal{E}}_{2}^r  \\
	&+& e^{-\ii k_z z} A_{i}^{r} \bm{\mathcal{E}}_{3}^r + e^{-\ii k_z z} B_{i}^{r} \bm{\mathcal{E}}_{4}^r, \nonumber \\
	\mathbf{H}_r &=& e^{\ii k_z z} A_{o}^{r} \bm{\mathcal{H}}_{1}^r + e^{\ii k_z z} B_{o}^{r} \bm{\mathcal{H}}_{2}^r  \\
	&+& e^{-\ii k_z z} A_{i}^{r} \bm{\mathcal{H}}_{3}^r + e^{-\ii k_z z} B_{i}^{r} \bm{\mathcal{H}}_{4}^r, 
\end{eqnarray*}
where the subscript $i(o)$ denotes incoming (outgoing) waves on the boundary, and $\kappa_n$, $n=1,2,3,4$ are solutions of \eqref{eq:roots}.  We have $8$ amplitudes, $A_{i,o}^{l,r}, B_{i,o}^{l,r}$ to be defined.  They are related by the scattering matrix which is defined through boundary (matching) conditions.

The \textit{in} and \textit{out} states and $\mathbf{S}$-matrix read
\begin{equation}
	\mathbf{E}^{out} = 
	\begin{pmatrix}
		A^l_o \\
		B^l_o \\
		A^r_o \\
		B^r_o 
	\end{pmatrix}, \ 
	\mathbf{E}^{in} =
	\begin{pmatrix}
		A^l_i \\
		B^l_i \\
		A^r_i \\
		B^r_i  
	\end{pmatrix},\ 
	\mathbf{E}^{out}  = 
	\mathbf{S} \cdot 
	\mathbf{E}^{in},
\end{equation}
where 
\begin{equation}
	\mathbf{S} = 
	\begin{pmatrix}
		\rb & \tb' \\
		\tb & \rb'
	\end{pmatrix},\ \rb = 
	\begin{pmatrix}
		r_{xx} & r_{xy} \\
		r_{yx} & r_{yy}
	\end{pmatrix}, \ 
	\tb = 
	\begin{pmatrix}
		t_{xx} & t_{xy} \\
		t_{yx} & t_{yy}
	\end{pmatrix}.
\end{equation}

To obtain $\mathbf{S}$-matrix we use the boundary conditions
\begin{eqnarray}
	\left.(\mathbf{E}^l - \mathbf{E}^r)\times \bm{n}_{l\to r}\right|_{z=0} &=&\bm{0},\nonumber \\
	\left.(\mathbf{H}^l - \mathbf{H}^r)\times \bm{n}_{l\to r} \right|_{z=0} &=& 4\pi\left.\bm{\sigma}_s \mathbf{E}^r\right|_{z=0},\label{eq:boundary}
\end{eqnarray}
where $\bm{\sigma}_s$ is (possible) tensor conductivity on the boundary $z=0$. 

We use the boundary conditions \eqref{eq:boundary}. Because $\bm{n} = (0,0,1)$ we may rewrite these equations in the following form
\begin{equation}
\eb^l = \eb^r,\ \hb^l = \hb^r + \frac{4\pi}{c} \left(\widetilde{\bm{\sigma}}_s \eb^r \right),
\end{equation} 
where $\eb, \hb$ are from Eq. \eqref{e,h} and
\begin{equation}
\widetilde{\bm{\sigma}}_s = 
\begin{pmatrix}
-\sigma_{21} & -\sigma_{22}\\
\sigma_{11} & \sigma_{12}
\end{pmatrix}.
\end{equation}

Taking into account Eqs. \eqref{eq:vv} and \eqref{eq:vnv}, the boundary conditions are represented in the following form   
\begin{equation*}
A_o^l \vb_3^l + B_o^l \vb_4^l - A_o^r \vbh_1^r - B_o^r \vbh_2^r  = \!  - A_i^l \vb_1^l \! - B_i^l \vb_2^l   + A_i^r \vbh_3^r + B_i^r \vbh_4^r.
\end{equation*}
By solving these relations we obtain components of the $\mathbf{S}$-matrix
\begin{eqnarray}
\rb &=& 
-\frac{1}{\Delta}\begin{pmatrix}
\begin{vmatrix}
\vb_1^l & \vb_4^l & \vbh_1^r & \vbh_2^r
\end{vmatrix} 
& 
\begin{vmatrix}
\vb_2^l & \vb_4^l & \vbh_1^r & \vbh_2^r
\end{vmatrix} \\[1ex]
\begin{vmatrix}
\vb_3^l & \vb_1^l & \vbh_1^r & \vbh_2^r
\end{vmatrix}
& 
\begin{vmatrix}
\vb_3^l & \vb_2^l & \vbh_1^r & \vbh_2^r
\end{vmatrix}
\end{pmatrix},\nonumber \\ 
\tb' &=& 
+\frac{1}{\Delta}\begin{pmatrix}
\begin{vmatrix}
\vbh_3^r & \vb_4^l & \vbh_1^r & \vbh_2^r
\end{vmatrix}
& 
\begin{vmatrix}
\vbh_4^r & \vb_4^l & \vbh_1^r & \vbh_2^r
\end{vmatrix}\\[1ex]
\begin{vmatrix}
\vb_3^l & \vbh_3^r & \vbh_1^r & \vbh_2^r
\end{vmatrix}
&
\begin{vmatrix}
\vb_3^l & \vbh_4^r & \vbh_1^r & \vbh_2^r
\end{vmatrix}
\end{pmatrix},\nonumber \\
\tb &=& 
+\frac{1}{\Delta}\begin{pmatrix}
\begin{vmatrix}
\vb_3^l & \vb_4^l & \vb_1^l & \vbh_2^r
\end{vmatrix}
& 
\begin{vmatrix}
\vb_3^l & \vb_4^l & \vb_2^l & \vbh_2^r
\end{vmatrix}\\[1ex]
\begin{vmatrix}
\vb_3^l & \vb_4^l & \vbh_1^r & \vb_1^l
\end{vmatrix}
&
\begin{vmatrix}
\vb_3^l & \vb_4^l & \vbh_1^r & \vb_2^l
\end{vmatrix} 
\end{pmatrix},\nonumber \\ 
\rb' &=& 
-\frac{1}{\Delta}\begin{pmatrix}
\begin{vmatrix}
\vb_3^l & \vb_4^l & \vbh_3^r & \vbh_2^r
\end{vmatrix}
& 
\begin{vmatrix}
\vb_3^l & \vb_4^l & \vbh_4^r & \vbh_2^r
\end{vmatrix}\\[1ex]
\begin{vmatrix}
\vb_3^l & \vb_4^l & \vbh_1^r & \vbh_3^r
\end{vmatrix}
&
\begin{vmatrix}
\vb_3^l & \vb_4^l & \vbh_1^r & \vbh_4^r
\end{vmatrix}
\end{pmatrix}, \label{eq:Smatrix}
\end{eqnarray}
where
\begin{equation*}
\Delta = |\vb_3^l \ \vb_4^l \ \vbh_1^r \ \vbh_2^r|, \ 
\vbh_n^r = \vb_n^r + \vb_n^\sigma,\ 
\vb_n^\sigma =
\begin{pmatrix}
\bm{0}\\
\frac{4\pi}{c}\widetilde{ \bm{\sigma}}_s \eb_n^r
\end{pmatrix},
\end{equation*}
and the vertical lines mean determinant. 

\bigskip
\textbf{III Rarefication procedure}

To rarefy matter we use relation $\varepsilon_{\mu\nu} = \delta_{\mu\nu} + 4\pi N \alpha_{\mu\nu}$ and expand expression for $\rb'$ up to first order in $N$. After long calculations we obtain 
\begin{eqnarray*}
&&r'_{ij} = \frac{\pi N}{k_z^2} (\alpha_{33} k_i k_j - \omega^2 \alpha_{ij} + k_i \alpha_{nj}k^n)\\
+&& \frac{\pi N}{k_z^3} (k_ik_j \alpha_{n3}k^n -  \kb ^2\alpha_{i3} k_j + k_z^2 (k_i \alpha_{3j} - \alpha_{i3} k_j)).
\end{eqnarray*} 
Because of 
\begin{equation}
\ln\det (\Ib - \epsilon \mathbf{A}) = 1 - \epsilon \tr\mathbf{A} + O(\epsilon^2),
\end{equation}
then from \eqref{eq:CaMat} we obtain 
\begin{equation}
\mathcal{E}_\Ca =  - \iint \frac{d^2k}{2(2\pi)^3} \int_{-\infty}^\infty d\xi e^{-2 a \kappa} \tr(\rb' \rb) + O(N^2).
\end{equation}
Straightforward calculation gives
\begin{eqnarray*}
\tr (\rb' \rb) &=& -\frac{\pi N}{\kappa^2} r^{ij}\left\{\xi^2 \alpha_{ji} +  \alpha_{ni} k^n k_j + \alpha_{33} k_ik_j   \right\} \\
&-& \sgn(\xi) \frac{\ii \pi N }{\kappa^3} r^{ij} \left\{ \kappa^2 [\alpha_{3i} k_j - \alpha_{j3} k_i]\right. \\
&-& \left. k_i ( \kb ^2\alpha_{j3} - \alpha_{n3}k^n k_j) \right\}.
\end{eqnarray*}
The last imaginary term gives no contribution to the energy and we arrive with expression
\begin{eqnarray*}
\mathcal{E}_\CP &=& \iint \frac{d^2k}{(2\pi)^2} \int_{0}^{\infty}\!\! \frac{d\xi}{\kappa}  e^{-2 \kappa  a} r^{ij} \\
&\times&\left[\xi^2 \alpha_{ji} +  \alpha_{ni} k^n k_j + \alpha_{33} k_ik_j\right].  
\end{eqnarray*}

%

\end{document}